\documentclass[aps,prb,showpacs,twocolumn,10]{revtex4-1}
\pdfpageattr {/Group << /S /Transparency /I true /CS /DeviceRGB>>}
\usepackage{graphicx}
\usepackage{amsmath}
\usepackage{bm}
\usepackage{hyperref}
\hypersetup{%
pdftitle={Scattering of two-dimensional massless Dirac electrons by a circular potential barrier},%
pdfauthor={J.-S. Wu et al.},%
pdfpagemode={UseNone},%
pdfstartview={FitH},%
breaklinks=true,%
citecolor=blue,%
colorlinks=true,%
linkcolor=blue,%
urlcolor=blue}
\usepackage{multirow}
\usepackage{xcolor}

\usepackage{CJKutf8}	

\begin{document}
\begin{CJK*}{UTF8}{bkai}	

\title{Scattering of two-dimensional massless Dirac electrons by a circular potential barrier}

\author{Jhih-Sheng Wu (\CJKchar{"54}{"33}\CJKchar{"81}{"F4}\CJKchar{"76}{"DB})
 and Michael M. Fogler}
 \affiliation{University of California San Diego, 9500 Gilman Drive, La Jolla, California 92093, USA}


\date{\today}
\begin{abstract}

We calculate the differential, total, and transport cross-sections for
scattering of two-dimensional massless Dirac electrons by a circular barrier.
For scatterer of a small radius, the cross-sections are dominated by quantum effects such as
resonant scattering that can be computed using the partial-wave series.
Scattering by larger size barriers is better described within the classical picture of reflection and refraction of rays,
which leads to phenomena of caustics, rainbow, and critical scattering.
Refraction can be negative if the potential of the scatterer is repulsive,
 so that a $p$-$n$ junction forms at its
boundary.
Qualitative differences of this case from the $n$-$N$ doping case are examined.
Quantum interference effects beyond the classical ray picture are also considered, such as normal and anomalous diffraction,
and also whispering-gallery resonances. 
Implications of these results for transport and scanned-probe experiments in graphene and topological insulators are  discussed.

\end{abstract}


\maketitle
\end{CJK*} 

\section{Introduction}
\label{sec:intro}

Recently, much interest has been attracted by electronic properties of materials where quasiparticles
 behave as massless two-dimensional (2D) Dirac fermions. Examples of such materials include
 graphene~\cite{Castro2009} and surface states of topological insulators.~\cite{Fu2007,Hasan2010}
Graphene has been studied more extensively because of advances in sample fabrication and a number
 of exceptional virtues, including a wide tunability of doping level and superior transport properties.
The latter are characterized by mean-free paths approaching several microns and the corresponding
 transport times $\tau_{\mathrm{tr}}$ in the range of picoseconds.~\cite{Castro2009, Peres2010ctp, DasSarma2011eti, Basov2014cgs}
However, scattering mechanisms limiting the transport mobility of graphene are still not fully understood.
For instance, a weak dependence of the mobility on the impurity density found in some experiments~\cite{Schedin2007}
remains an open problem. It was argued that this weak dependence could be due to
correlations in impurity positions. Both negative, i.e., repulsive\cite{Li2011} and positive, i.e., attractive
correlations could impact the mobility. An example of the latter is aggregation of impurities into clusters of
size of tens of nanometers. Modelled as circularly symmetric potential barriers with sharp boundaries,
such finite-size scatterers were predicted~\cite{Katsnelson2009} to degrade the mobility much less compared
to random uncorrelated impurities. Another observable signature of finite-size scatterers is a significant difference
between the transport time $\tau_{\mathrm{tr}}$ and the quantum lifetime $\tau_q$.
(The latter can be extracted from magnetotransport measurements.) Although the ratio
\begin{equation}
\eta \equiv \tau_{\mathrm{tr}} / \tau_q
\label{eq:eta_def}
\end{equation}
varies widely among different experiments, it can be as high as a factor of six.~\cite{Gorbachev2009}
For massive 2D fermions scattered by random sharp barriers,
$\eta$ should approach $3 / 2$ when the barriers become impenetrable.~\cite{Yudson2007}
Since graphene quasiparticles behave as massless fermions,
they can penetrate arbitrary high potential barriers (lower than the total energy bandwidth) via
the process of Klein's tunneling and associated
negative refraction.~\cite{Katsnelson2006cta, Cheianov2007, Cserti2007}
Therefore, the dependence of $\eta$ on the barrier parameters
is another open problem.
For these and other reasons, scattering of quasiparticles by finite-size defects warrants further qualitative and quantitative investigation
in order to better understand transport and magnetotransport properties of graphene.
While there have been already a number of previous studies of circular potential barriers in graphene,~\cite{Katsnelson2007, Cserti2007, Guinea2008, Matulis2008, Hewageegana2008, Bardarson2009,Katsnelson2009, Heinisch2013}
they have not elucidated in a comprehensive way how scattering by such a barrier depends on its size and strength.
Furthermore, some of this prior work contains minor errors.
In this paper, we correct, refine, and extend these investigations.

The question of what limits the surface electron conduction in topological insulators is even more wide open.
There are few studies that examined scattering of Dirac fermions by circular potential barriers in this context
although other types of scattering defects have been considered.~\cite{Parente2011, Fu2011, Parente2014, Masir2011a, Masir2011b, Masir2013}
Most of our results for this problem should also apply to Dirac fermions at the surface of topological insulators.

\begin{figure}[b]
\includegraphics{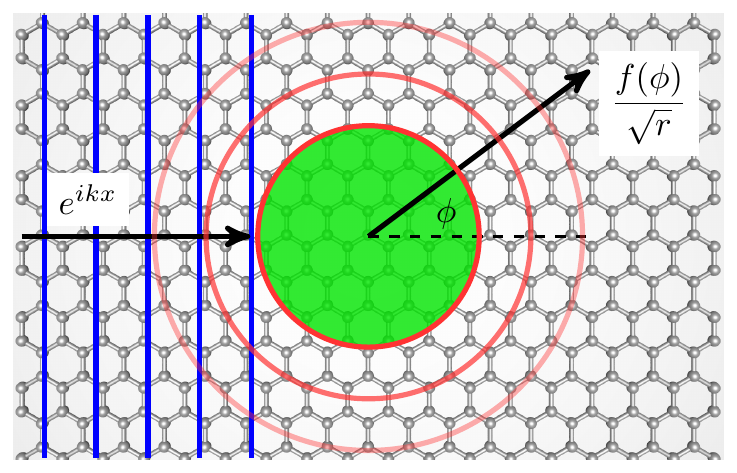}
\caption{(Color online) Electron plane wave scattered by a circularly symmetric step-like potential in a Dirac metal such as graphene or a surface of a topological insulator.
\label{fig:model}
}
\end{figure}

The effective low-energy Hamiltonian of the model we study is~\cite{Katsnelson2007, Cserti2007, Guinea2008, Matulis2008, Hewageegana2008, Bardarson2009,Katsnelson2009, Heinisch2013}
\begin{equation}
{H} = v_{F}\, (\sigma_{x} p_x + \sigma_{y} p_y) + V(r)\,,
\label{eq:dirac_hamil}
\end{equation}
where $p_\nu$ are the momentum operators, $\sigma_{\nu}$ are the Pauli matrices, and $v_{F} \sim 10^{8}\,\mathrm{cm}/\mathrm{s}$ is the Fermi velocity.
The potential $V(r)$ is assumed to be step-like,
\begin{equation}
V(r) = V_{0}\, \theta(a-r),\label{eq:potential}
\end{equation}
where $\theta(r)$ is the unit step function and $a$ is the radius of the disk. 
The scattering of an electron with energy $E>0$ by this potential is characterized by two dimensionless parameters,
\begin{align}
 X &=\dfrac{E a}{\hbar v_{F}} \equiv k a
 \quad\textrm{and}\quad
 \rho = -\dfrac{V_0 a}{\hbar v_{F}}\,,
\end{align}
which specify the size and the strength of the barrier, respectively;
$X$ also gives an estimate of the maximum angular momentum involved in the scattering.

\begin{figure}[t]
\includegraphics{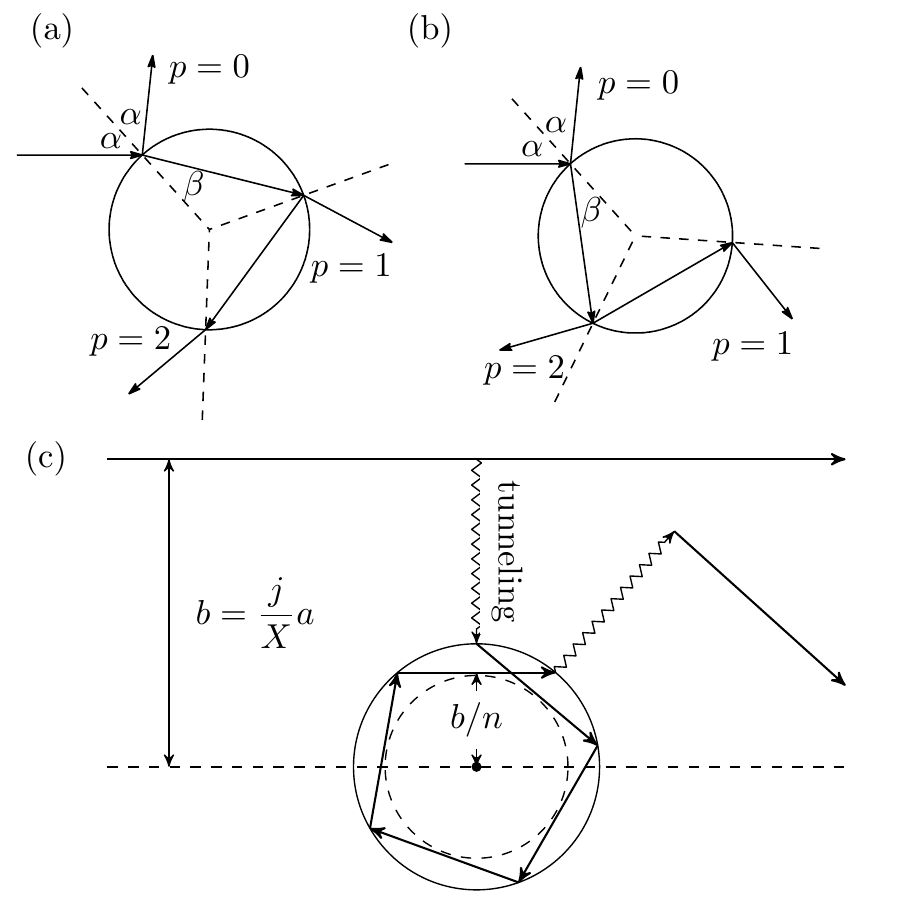};
\caption{(a) The first three rays $p = 0$, $1$, $2$ in the ray series for an $n$-$N$ junction (b) 
Same for an $n$-$p$ junction.
(c) Ray tunneling and a whispering gallery mode inside the scatterer. 
} \label{fig:ray_pic}        
\end{figure} 

Consider a plane wave incident on the scatterer.
The problem is to find the far-field scattering amplitude $f(\phi)$ as a function of the deflection angle $\phi$, see Fig.~\ref{fig:model}. 
The differential cross-section is then calculated from
\begin{align}
\dfrac{d\sigma}{d\phi}=|f(\phi)|^2\,.
\end{align}
The formal solution of this problem is given by the standard partial-wave decomposition (PWD)
\begin{align}
f(\phi)=-\dfrac{i}{\sqrt{2\pi k}}\sum_{j}\left(e^{2i\delta_{j}}-1\right)
          e^{i(j-1/2)\phi},
\label{outside_f}    
\end{align}
where $\delta_{j}$ is the phase shift for angular momentum $j$.
The well-known peculiarity of Dirac fermions is that $j$'s are not integers but half-integers, see 
 Appendix~\ref{subsec:pwd}.

Integrating the differential cross-section over $\phi$, we can express
the total and the transport cross-sections in terms of the phase shifts~\cite{Shytov2007}
\begin{align}
\sigma&=\dfrac{4}{k}\sum_{j}\sin^{2}\delta_{j}
\label{eq:crs_1}\,,\\
\sigma_{\mathrm{tr}}&=\dfrac{4}{k}\sum_{j>0}\sin^{2}(\delta_{j+1}-\delta_{j})\,.
\label{eq:tcrs_1}
\end{align} 
The total cross-section obeys the optical theorem
\begin{align}
\sigma=\sqrt{\dfrac{8\pi}{k}}\, \textrm{Im}f(0)\,.\label{eq:op_theorem}
\end{align}
For $X \ll 1$ the PWD is dominated by a first few terms and the result has an intuitive interpretation in terms of resonant scattering.
Conversely, 
for $X\gg1$, PWD suffers from slow convergence and lacks a transparent physical meaning. 
In fact, it appears that numerical results reported in a previous work on this problem~\cite{Katsnelson2009} are inaccurate
[See the second paragraph below Eq.~\eqref{eq:res_3}.]
The main effort in this paper is devoted to treating this difficult $X \gg 1$ regime by alternative methods.
Especially instructive one is the semiclassical approximation.
This approach leads to the so-called 
ray series, which accounts for most of the observable features of the $X \gg 1$ regime and has an intuitive representation in terms of ray paths (Fig.~\ref{fig:ray_pic}).
In this context it is convenient to introduce another dimensionless parameter, the refractive index:
\begin{align}
n = \dfrac{X'}{X} = 1 + \dfrac{\rho}{X}\,,
\label{eq:n}
\end{align}
where $X'$ is defined as
\begin{align}
 \quad X' \equiv X + \rho\,.
\end{align}
The refractive index $n$ can be of either sign.
If it is positive (negative), we deal with, respectively, $n$--$N$ and $n$-$p$ junction at $r = a$.
In the latter case, realized for $\rho < - X$, the Dirac quasiparticles exhibit the aforementioned negative refraction.\cite{Cheianov2007}
This modifies the ray trajectories qualitatively,\cite{Cserti2007} cf.
Figs.~\ref{fig:ray_pic}(a) and \ref{fig:ray_pic}(b) and Sec.~\ref{sec:ray_pic}.

\begin{figure}[b]
\includegraphics{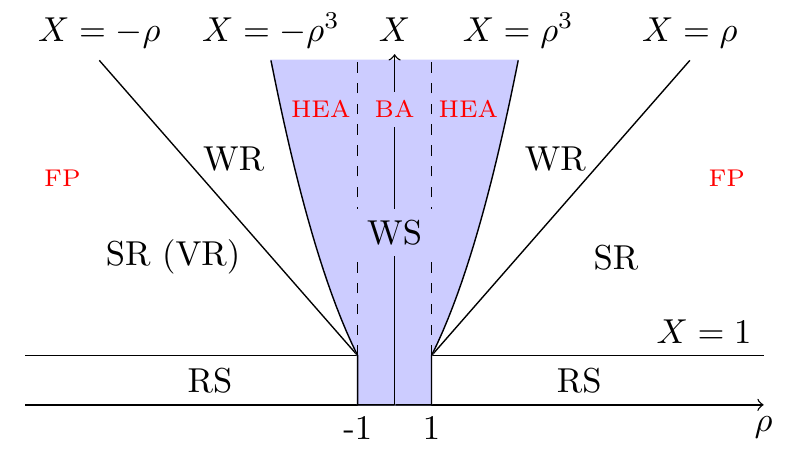}
\caption{(Color online) Regime diagram of the scattering. 
RS: resonant scatterer, SR: strong reflector,
FP: Fabry-P\'erot resonator,
WR: weak reflector, WS: weak scatterer, HEA: high energy approximation,
BA: Born approximation. VR: Veselago reflector, where  negative refraction occurs.
}\label{fig:regime}
\end{figure}

The remainder of the paper is organized as follows. 
In Sec.~\ref{sec:main_results} we classify the regimes of scattering according to $X$ and $\rho$.
We present the global regime diagram (Fig.~\ref{fig:regime}) and give the formulas for the cross-sections in each regime.
We find the most feature-rich case to be $X \gg 1$.
In Sec.~\ref{sec:ray_pic}, we study this case using the semiclassical method.  
In Sec.~\ref{sec:sws}, we discuss phenomena beyond the semiclassical approximation.
In Sec.~\ref{sec:feature_dcrs}, the angular dependence of the differential cross-section in various large-$X$ regimes is analyzed.
Besides far-field scattering, we also consider the structure of the electron wavefunctions near the scatterer.
In Sec.~\ref{sec:nfs}, we briefly discuss implications of these near-field effects for scanned-probe experiments with graphene and topological insulators. 
In Sec.~\ref{sec:conclusion}, we summarize our contributions and comment on 
possible future extensions of our study.
The general outline of our analytical derivations is presented in Secs.~\ref{sec:ray_pic}--\ref{sec:feature_dcrs}, with the
additional details provided in the Appendices.

\section{Main results}
\label{sec:main_results}

We start with a brief overview of different scattering regimes indicated in Fig.~\ref{fig:regime}.
These regimes are classified according to the behavior of the cross-sections as functions of $X$ and $\rho$. 
The horizontal line $X = 1$ partitions Fig.~\ref{fig:regime} into two domains.
In the upper one, $X \gg 1$, the scattering has a predominantly
semiclassical character.
Except for the region of small deflection angles $\phi<1/X$, which is governed by diffraction,
the scattering amplitude is obtained by summing the ray series  (Fig.~\ref{fig:ray_pic}), expressed mathematically by the Debye expansion.~\cite{Nussenzveig1969b, grandy2000} 
In the lower domain, $X \ll 1$, scattering is dictated by quantum effects and the ray picture generally does not apply.

In the strong reflector (SR) and weak reflector (WR) regimes of
Fig.~\ref{fig:regime}, the interference between the rays and the diffraction can be neglected.
As a consequence of the Babinet principle,
each of the two contributes $2a$ to the total cross-section,\cite{Yudson2007}
and so
the total cross-section is
\begin{align}
{\sigma} \simeq 4{a}
\quad\left(\textrm{SR and WR regimes}\right)\,.
\label{eq:4a}
\end{align} 
In contrast, in the weak scatterer (WS) regime of Fig.~\ref{fig:regime},
where most of the rays are scattered by small angles,
the interference of the ray and diffraction amplitudes becomes important.
In this ``anomalous diffraction" (AD) regime~\cite{van_de_Hulst1957lss}
the ray picture fails. Instead,
the scattering can be dealt with the perturbation theory,
such as the high energy approximation (HEA) and
the Born approximation (BA).
As discussed below, Eq.~\eqref{eq:4a} becomes replaced by more complicated expressions, Eqs.~\eqref{eq:crs_HEA} and \eqref{eq:crs_BA},
that predict oscillations of $\sigma$ as a function of $\rho$.
The BA in fact describes the entire $|\rho| \ll 1$ strip in Fig.~\ref{fig:regime}, including the $X \ll 1$ part.

\begin{figure}
\includegraphics{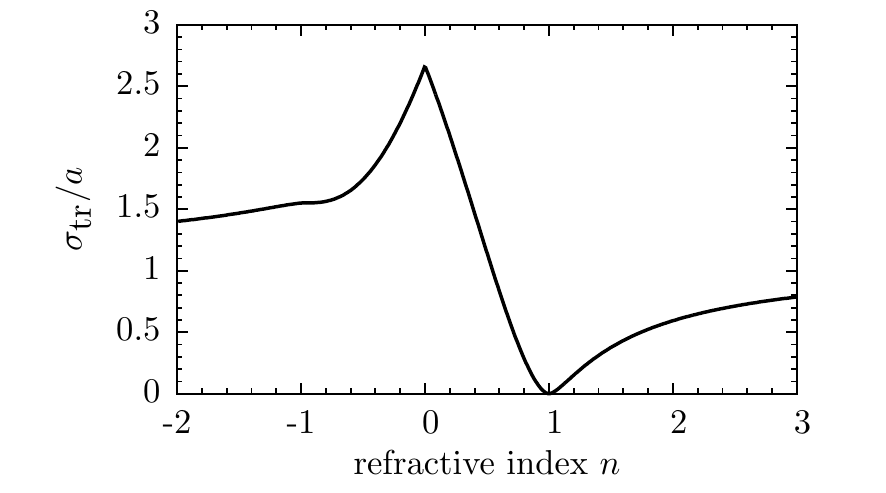}
\caption{Analytical approximation [Eq.~\eqref{eq:tcrs_ray}] for the transport cross-section as a function of the refractive index.}\label{fig:tcrs_ana}
\end{figure}

In the rest of the quantum domain $X \ll 1$, $|\rho| \gg 1$, the cross-sections ${\sigma}$ and $\sigma_{\textrm{tr}}$ are determined by resonant scattering (RS), see Fig.~\ref{fig:regime}.
These cross-sections can be efficiently computed by summing the partial-wave series, Eqs.~\eqref{eq:crs_1} and \eqref{eq:tcrs_1}.
Unlike the $|\rho| \ll 1$ case, where
the lowest angular momenta~\cite{Katsnelson2007} $j = \pm 1/2$ dominate,
here significant contributions arise from certain high 
$j$ for which the resonant tunneling condition is satisfied.
 
Let us now give more detailed information about each of the regimes.
The transport cross-sections in the SR and WR regimes are dominated by the ray series, since the diffraction is restricted to small angles.
Neglecting interference among different rays,
we obtained the result
\begin{align}
\dfrac{\sigma_{\textrm{tr}}}{a} &= \dfrac{8}{3}
- \textrm{sgn}(n) \varsigma(n)\,,
\label{eq:tcrs_ray}\\
\varsigma(n) &= 4\!\!\! \int\limits_{0}^{\textrm{min}(1,|n|)}\!\!\!
d b
\dfrac{(2n-1)b^2-n}{n^2-(2n-1)b^2}\sqrt{n^2-b^2}\sqrt{1-b^2}
\label{eq:d}\\
&(\textrm{SR and WR regimes}).\notag
\end{align}
The correction term $\varsigma(n)$ can also be written as a linear combination of the complete elliptical integrals, see Sec.~\ref{sec:ray_pic}.
Equation~\eqref{eq:tcrs_ray} predicts the following asymptotic behavior of the transport cross-section:
\begin{align}
\dfrac{ \sigma_{\textrm{tr}} }{a} \simeq
&\left\{\begin{array}{cc}
3(n-1)^2\ln\left|\dfrac{1}{n-1}\right|, & n \simeq 1\,,\\[.7em]
 \dfrac{8}{3}- 2 \pi (\sqrt{2}\, - 1) |n|, & |n|\ll 1\,,\\[.7em]
\dfrac{8}{3}-\dfrac{\pi}{2}, & |n| \gg 1
\end{array}\right.&
\label{eq:tcrs_ray_limit}\\[0.6em]
&(\textrm{SR and WR regimes}).&\notag
\end{align}
Notable features include the zero of $\sigma_{\textrm{tr}}$ at $n=1$ and the plateau-like inflection point at $n = -0.96$ (see Fig.~\ref{fig:tcrs_ana}).
The left SR regime in Fig.~\ref{fig:regime} corresponds to the negative refractive index $n < 0$, and so we gave it an additional appellation of ``Veselago reflector'' (VR).
The typical ray trajectories are shown in Fig.~\ref{fig:ray_pic}(b). 
The left diagonal line $X = -\rho$ separating the WR and the SR(VR) regimes corresponds to $n = 0$. Along this line the ray formalism predicts a cusp in $\sigma_{\textrm{tr}}$, see Fig.~\ref{fig:tcrs_ana} and the second line of Eq.~\eqref{eq:tcrs_ray_limit}.
(In reality, the cusp is rounded and shifted by $\mathcal{O}(1 / X)$ quantum corrections, cf.~Fig.~\ref{fig:katz} below.)
On the other hand, transport cross-section varies smoothly
across the right diagonal line $X = \rho$ (or $n = 2$) in Fig.~\ref{fig:regime} separating the WR and the SR regimes.

\begin{figure}
\includegraphics[scale=1]{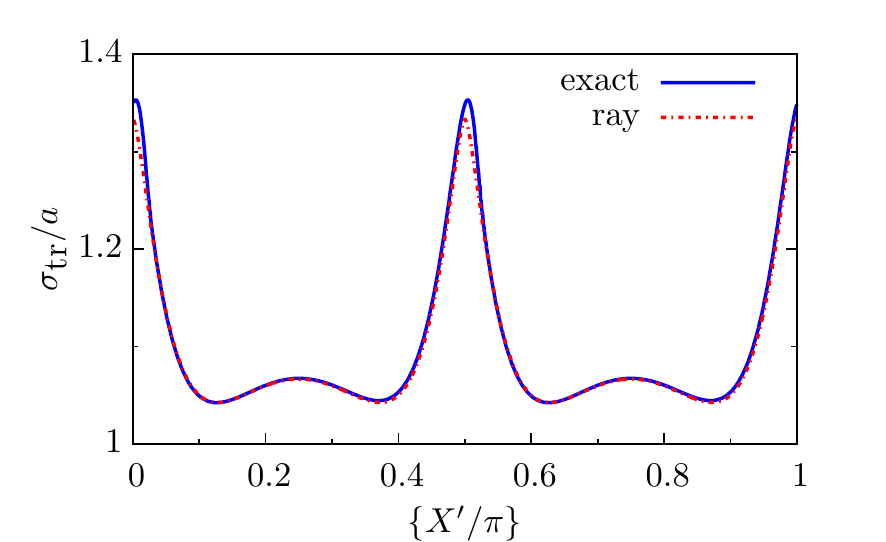}
\caption{(Color online) Transport cross-section for $|n|\gg 1$ case as a function of $\{X' / \pi\}$, the fractional part of $X' / \pi$.
The line labelled ``exact'' represents Eq.~\eqref{eq:tcrs_1}
evaluated for a fixed $X = 10^3$ and $\{X' / \pi\} = X' / \pi - 10^7$.
The line labelled ``ray'' is Eq.~\eqref{eq:tcrs_inf_n}. 
}\label{data_inf_n}
\end{figure}

Another analytical result can be derived in the limit $|n|\rightarrow\infty$, which describes the leftmost and rightmost parts of the SR regimes in Fig.~\ref{fig:regime}.
The rays pass almost through the center of the disk in this limit, and so 
it is possible to sum the ray amplitudes fully taking into account their interference and obtain
\begin{align}
\dfrac{\sigma_{\textrm{tr}}}{a} &\simeq \dfrac{8}{3}+\sec^5 2 X' \left[\cos 2 X' - \dfrac{7}{3} \cos 6 X'\right.\notag\\
&+\left.\dfrac{1}{4}\left(-8\cos 4 X' + \cos 8 X' + 7\right)
\ln \tan^2 X'\right]
\label{eq:tcrs_inf_n}\\[0.6em]
& (\textrm{SR regime,}\ |n|\rightarrow\infty \textrm{ limit}).\notag
\end{align}
This expression is $\pi / 2$-periodic in $X'$ and if $\rho$ is fixed, also in $X$, as expected for the Fabry-P\'erot (FP) resonator of length $2a$. Figure~\ref{data_inf_n} shows  the transport cross-section and
the comparison between the exact result and Eq.~\eqref{eq:tcrs_inf_n}.

The border of the WS regime in Fig.~\ref{fig:regime} is defined by the curves $X^3=\pm\rho$ at $X > 1$ and $\rho = \pm 1$ at $X < 1$.
The $X > 1$ part is described by the HEA.
More precisely, the HEA describes the small-angle part of the scattering amplitude; however,
it is sufficient for computing both the total and transport cross-sections in the WS regime.
The total cross-section is given by
\begin{subequations}
\begin{align}
\dfrac{\sigma}{a} =\ & 2\pi  \textrm{H}_{1}(2\rho)
\label{eq:crs_HEA}\\
 \simeq\
 &\begin{cases}
        \dfrac{16}{3}\,\rho^2, & \rho \ll 1\,,\\[.5em]
        4 \left[1 - \dfrac{\pi}{4\rho}
                   \cos\left(2\rho-\dfrac{\pi}{4}\right)
         \right],
        & \rho \gg 1
  \end{cases}&\\[0.6em]
&(\textrm{HEA regime}),&\notag
\end{align}
\end{subequations}
where $\textrm{H}_{1}(z)$ is the Struve function.
The transport cross-section has the form
\begin{align}
\dfrac{\sigma_{\textrm{tr}}}{a} \simeq \dfrac{2\rho^2}{X^2}\,
\ln X
\quad ({\textrm{HEA regime}}).
\label{eq:tcrs_HEA}
\end{align}
The $|\rho| \ll 1$ part of the HEA domain is alternatively described
by the BA.
Additionally, the BA gives the form of the differential cross-section for arbitrary angles $\phi$:
\begin{align}
\dfrac{d\sigma}{d\phi} =\dfrac{\pi}{2}\,
\dfrac{\rho^2 a}{X}\,
\cot^2 \left(\dfrac{\phi}{2}\right)
J_{1}^2 \left(2 X\sin\dfrac{\phi}{2}\right)
\quad \textrm{(BA)}.
\label{df_BA}
\end{align} 
Integrating Eq.~\eqref{df_BA} over $\phi$, we obtain
\begin{align}
\dfrac{\sigma}{a} \simeq
&\begin{cases}
\dfrac{16}{3}\,\rho^2, & X\gg1\,,\\[.6em]
\dfrac{\pi^2}{2}\, \rho^2 X , & X \ll 1\,,
\end{cases}&
\label{eq:crs_BA}\\[.6em]
\dfrac{\sigma_{\textrm{tr}}}{a} \simeq
&\begin{cases} \dfrac{2\rho^2}{X^2}\, \ln X, & X \gg 1\,, \\[.6em]
\dfrac{\pi^2}{4}\, \rho^2 X, & X \ll 1
\end{cases}&
\label{eq:tcrs_BA}\\[0.6em]
&\textrm{(BA regime).}&\notag
\end{align}
The upper lines in these equations agree with the HEA.
The formulas on the lower lines differ from Eq.~(12) of Ref.~\onlinecite{Guinea2008} and Eq.~(51) of Ref.~\onlinecite{Katsnelson2007} proposed earlier for the same regime.
We believe ours to be the correct ones. 
Unlike in the SR and WR domains,
the cross-sections in the WS regime (both HEA or BA) cannot be written solely in terms of
the classical quantities $n$ and $a$.

A component missing in the ray series
is the resonant tunneling of the rays with impact parameters $b$ larger than the radius of the disk [Fig.~\ref{fig:ray_pic}(c)].
These rays correspond to the partial wave with $j = k b > X$.
Within the semiclassical picture the region $a < r < b$ is classically forbidden due to the ``centrifugal'' potential barrier.
The tunneling through such a barrier is usually exponentially small unless the resonance condition is met.
For certain values of $\rho$ and $X$, tunneling of the waves with specific $\pm j_{r}$ becomes strongly enhanced, which creates sharp maxima of the cross-sections.\cite{grandy2000}
The resonant tunneling may be encountered for either type of doping.
A necessary condition for the resonance is that some of disk interior is classically allowed.
This is possible if $n > 1$ or $n < -1$, so that the interval $X < |j| < |X'|$ exists.
The condition for the $z$th resonance ($z = 1, 2, \ldots$)
can be derived from the Bohr-Sommerfeld quantization rule valid for $X \gg 1$.
This condition has the form
\begin{equation}
2\sqrt{X'^2-j^2}
- 2|j| \cos^{-1}\dfrac{|j| }{X'}
+ 2\Theta_+ - \dfrac{\pi}{2} = 2 \pi z\,,
\label{eq:whispering}
\end{equation} 
where $\Theta_+ \sim 1$ is the phase shift of the inner reflection at the disk boundary, cf.~Sec.~\ref{subsec:res}.
Since Eq.~\eqref{eq:whispering} is invariant under the sign change of $j$,
a pair $j = \pm j_r$ would resonate simultaneously.
From Eq.~\eqref{eq:crs_1} we see that each resonant partial wave with $|j| > X$ contributes up to $4 a / X$ to the cross-section, so each resonant pair contributes up to $8 a / X$.
This amount is parametrically small compared to the collective contributions $\sim a$ of all the $j < X$ partial waves. 
Hence, the resonances produce only a small ``ripple structure'' in the cross-section.\cite{Chylek1976} 
In contrast, at $X \ll 1$ and $\rho \gg 1$, the RS is the dominant effect.
In this regime, the cross-sections are given by the approximate formulas \cite{Hewageegana2008,Bardarson2009} (see Sec.~\ref{subsec:res})
\begin{align}
\dfrac{d\sigma}{d\phi}
  &\simeq \sum_{j \geq 1/2}
\frac{\sigma_j}{\pi} \cos^{2}j\phi\,,
\label{eq:res_1}\\
 \sigma_j &= \dfrac{8a}{X}\,
\sum_{z=1}^{\infty} \frac{\gamma^2}{(\rho - \rho_{j,z})^2 + \gamma^2}\,,
\label{eq:res_2}\\
 \sigma &\simeq \sum_{j\geq1/2}\sigma_j\,,
\quad
\sigma_{\textrm{tr}}
 \simeq \dfrac{\sigma_{1/2}}{2}+\sum_{j\geq3/2}\sigma_j
\label{eq:res_3}\\[0.6em]
&(\textrm{RS regime})\notag
\end{align}
with
$\rho_{j,z}$ and $\gamma$ defined by Eqs.~\eqref{eq:sin_delta_3} and \eqref{eq:gamma} in Sec.~\ref{sec:sws}.
The resonance widths $\gamma \sim {X^{2j}}$ and their $\rho$-integrated weights $\sim X^{2j - 1}$ rapidly decrease with $j$, which is consistent with the prevalent role of the lowest angular momentum resonance, $j = 1 /2$, for short-range scatterers.

\begin{figure}[b]
\begin{center}
\includegraphics[scale=1]{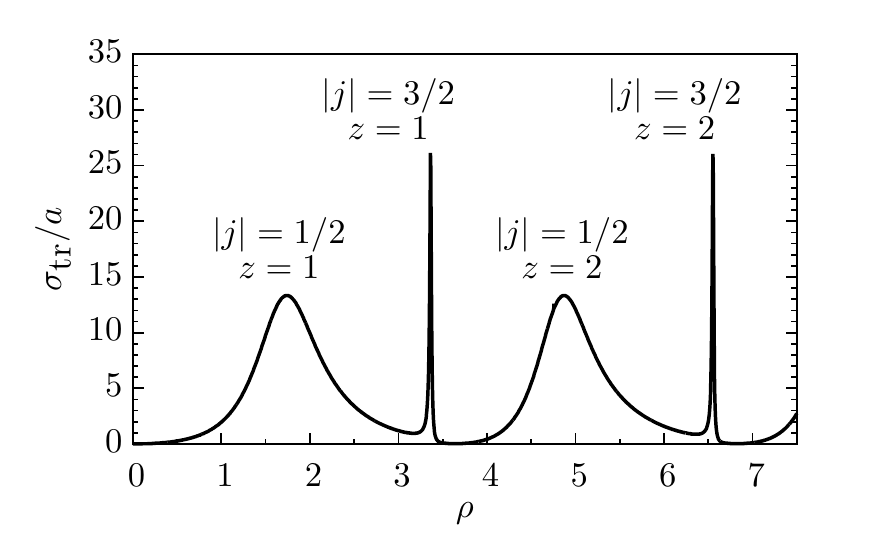}
\caption{Transport cross-section in the BA and RS regimes, $X = 0.3$.
Labels $j$ and $z$ denote the angular momentum of the resonant partial waves
and index of the quasibound state in the disk.
}
\label{fig:data_angle_dcrs}
\end{center}
\end{figure}  

Let us now illustrate some of the above formulas by specific examples.
The RS behavior is depicted in Fig.~\ref{fig:data_angle_dcrs},
which shows the differential cross-sections for $X = 0.3$ and varying $\rho$. Note that $j = 3 /2$ resonance is much more narrow that $j = 1 / 2$,
as stated above.
The BA behavior $\sigma_{\textrm{tr}} \propto \rho^2$ [Eq.~\eqref{eq:tcrs_BA}] is seen to occur at small $\rho$.

\begin{figure}
\includegraphics{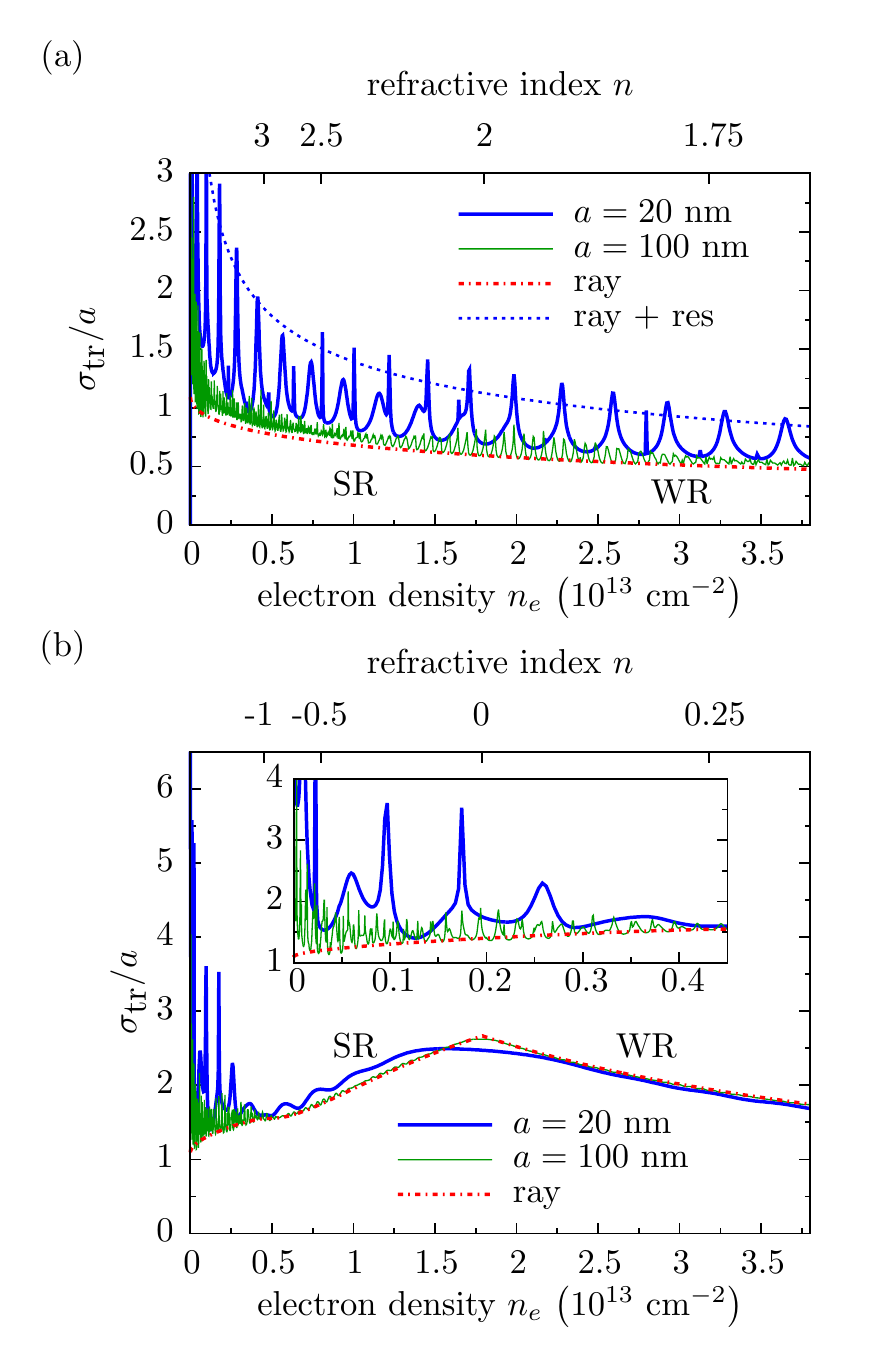}
\caption{(Color online) Transport cross-section as a function of electron density. (a) $V_0 = -0.5\,\mathrm{eV}$ or $\rho \approx +{300\,\textrm{nm}} / {a}$. (b) $V_0 = +0.5\,\mathrm{eV}$ or $\rho \approx -{300\,\textrm{nm}} / {a}$.
Thick solid curves: the exact result from PWD for $a = 20\,\mathrm{nm}$.
Thin solid curves: the exact result from PWD for $a = 100\,\mathrm{nm}$.
Dashed-dotted curves: Eq.~\eqref{eq:tcrs_ray}, from the ray picture.
Dotted curves: the sum of the ray term [Eq.~\eqref{eq:tcrs_ray}]
and a single resonance ${8a} / {X}$ for $a = 20\,\mathrm{nm}$.
The inset of (b) shows the effect of the resonances at small density where $n < -1$.
}\label{fig:katz}
\end{figure}  

Next, consider the dependence of the cross-section as
a function of the electron density, $n_e = k^2 /\pi = X^2 / (\pi a^2)$
for a fixed potential strength $V_0$ in SR and WR regimes.
Figure~\ref{fig:katz}(a) shows the transport cross-section for $V_0 = -0.5\,\mathrm{eV}$,
which models the effect of Al, Ag, or Cu adsorbates weakly coupled to graphene.\cite{Giovannetti2008}
In this case, the system is always a $n$-$N$ junction, i.e., $n > 0$.
The dashed-dotted curve in Fig.~\ref{fig:katz}(a) is given by the ray formula, Eq.~\eqref{eq:tcrs_ray}.
It fits well with the (numerically) exact results from PWD, especially for $a = 100\,\mathrm{nm}$.
For $a = 20\,\mathrm{nm}$, the deviations of the PWD curve from the ray formula are larger.
However, they almost never exceed $8a / X$, the
vertical shift between the dashed-dotted and dotted curves.
This shows that this contribution comes predominantly from a single resonant pair; bunching of the resonances is atypical.
We checked that the positions of the resonances are
rather well described by Eq.~\eqref{eq:whispering}.
Figure~\ref{fig:katz} also shows that as $n_e$ increases, the resonance contribution becomes smaller compared to the ray term, as expected because $X$ increases.
The curve marked $a = 20~\mathrm{nm}$ in Fig.~\ref{fig:katz}(a) is computed for the same parameters as in Fig.~2 of Ref.~\onlinecite{Katsnelson2009}. The results are clearly different in both the magnitude and the periodicity.
The reason for the disagreement is unknown because the same mathematical formula [Eq.~\eqref{eq:tcrs_1}] was used in our calculation and in Ref.~\onlinecite{Katsnelson2009}. We believe our numerical results are correct 
because they are consistent with our analytic formulas.

In Fig.~\ref{fig:katz}(b), we consider $V_0 = +0.5\,\mathrm{eV}$, where the system changes from 
an $n$-$p$ junction to a $n$-$N$ junction as the electron density increases. 
The ripple structure exists only at $n < -1$, in agreement with the condition discussed above.
It is worth noting that in the limit $|n| \ll 1$, the transport cross-section,  $\sigma_{\textrm{tr}}$ approaches the value of $8 a / 3$, the 
known result for an impenetrable disk.\cite{Yudson2007}
This is the maximum transport cross-section one can get for massless Dirac fermions at $X \gg 1$.
However, in the opposite limit $|n| \gg 1$, which would also correspond to impenetrable disk for massive fermions, Eq.~\eqref{eq:tcrs_ray_limit}
predicts a different and significantly smaller value $\sigma_{\textrm{tr}} = (8/3 - \pi/2) a \approx 1.1a$. This highlights the ability of massless Dirac to penetrate high barriers via Klein's tunneling.

\begin{figure}
\includegraphics{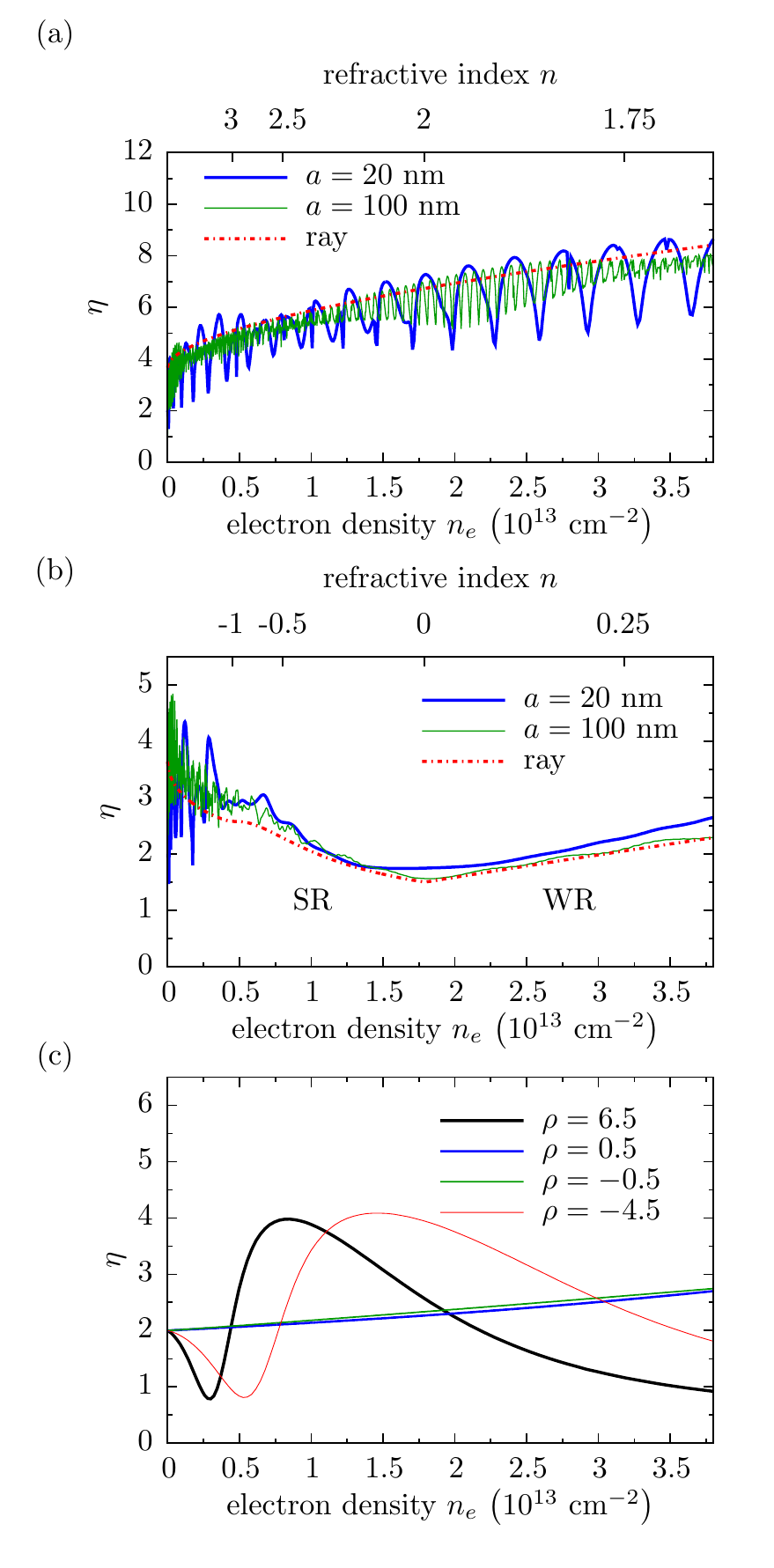}
\caption{(Color online) The ratio $\eta$ of the quantum and transport times as a function of the electron density $n_e$.
(a) $V_0 = -0.5\,\mathrm{eV}$ or $\rho \approx +{300\,\textrm{nm}} / {a}$,
same as in Fig.~\ref{fig:katz}(a). (b) $V_0 = +0.5\,\mathrm{eV}$ or $\rho \approx -{300\,\textrm{nm}} / {a}$,
same as in Fig.~\ref{fig:katz}(b).
(c) ``Point-like'' scatterer, $a = 1\,\mathrm{nm}$.}
\label{fig:lifetime} 
\end{figure} 

Consider now graphene with many randomly positioned identical disks of low enough
concentration $n_c \ll 1 / a^2$.
Our results can be used to compute the conductivity $G$
of such a system
if its size is larger than the mean-free path $l = v_F \tau_{\textrm{tr}}$.
The conductivity is related to the transport cross-section by
\begin{align}
G = \frac{2e^2}h{}k_F v_F \tau_{\textrm{tr}} = 2\sqrt{\pi}\,\frac{e^2}{h}\frac{\sqrt{n_e}}{n_c\sigma_{\textrm{tr}}}
\,.
\label{eq:conductance}
\end{align}
Note that Eq.~\eqref{eq:conductance} neglects logarithmic corrections due to multiple scattering effects.~\cite{Aleiner2006, Ostrovsky2006}
If the disks are slightly different in size or shape or if the system is at finite temperature,
we expect the ripple structure in $\sigma_{\textrm{tr}}$ to be washed out leaving only the overall trends.
For example, for parameters used in Fig.~\ref{fig:katz},
the disorder-averaged transport cross-section should change slowly with the carrier concentration $n_e$, remaining close to $\sim a$. 
Hence, the conductance will have an approximately $\sqrt{n_e}$ dependence.
Such a dependence is different 
from those for both the charged impurities and the short-ranged defects
computed within the BA in Ref.~\onlinecite{Hwang2008}, which are, respectively,
linear and constant in $n_e$.

An important parameter $\eta$ characterizing the spatial structure
of impurities is the ratio of the transport time $\tau_\textrm{tr}$ and the quantum lifetime $\tau_q$ [Eq.~\eqref{eq:eta_def}].
This parameter is related to the cross-sections via
\begin{align}
\eta=\dfrac{\tau_\textrm{tr}}{\tau_q}=\dfrac{\sigma}{\sigma_{\textrm{tr}}}. \label{eq:lifetime}
\end{align}
Experimentally, $\tau_{\textrm{tr}}$ can be extracted from the conductivity measured in the absence of magnetic field [Eq.~\eqref{eq:conductance}]
whereas $\tau_q$ can be obtained from the damping rate of the Shubnikov-de~Haas oscillations in magnetotransport.
A large $\eta$ indicates that scattering is predominantly in the
forward direction while a small $\eta$ indicates that
scattering by large angle is possible.
The former is a feature of long-range impurities.
The latter may indicate either that the impurities are short-range or they are have sharp boundaries.

In Fig.~\ref{fig:lifetime}(a) and Fig.~\ref{fig:lifetime}(b), we compute
$\eta$ for large scatterers and $V_0 = \pm0.5$~eV, respectively.
For the $n$-type scatterers ($V_0 < 0$) in Fig.~\ref{fig:lifetime}(a),
$\eta$ increases from $4$ to $7$ with the electron density,
which is a range of values found in the experiment.\cite{Gorbachev2009}
The ray formula is basically the envelope of the exact results. 
The ripple structure exists everywhere since $|n|>1$ for all the electron density.
For  the $p$-type scatterers ($V_0>0$) in Fig.~\ref{fig:lifetime}(b),
the ray formula fits even better except again for the ripple structure at $n < -1$.
Parameter $\eta$ exhibits a minimum at $n = 0$, which is due to the maximum of the transport cross-section at such $n$, Fig.~\ref{fig:katz}(b).

In Fig.~\ref{fig:lifetime}(c), we show $\eta$ computed for disks of small radius $a = 1\,\mathrm{nm}$.
Note that two scattering regimes are possible for small impurities, RS ($\rho \gg 1$) and WS ($\rho \ll 1$).
In the WS regime, the scattering dominated by the partial waves with $|j|=1/2$, so that $\eta \approx  2$, cf.~Eqs.~\eqref{eq:res_2} and \eqref{eq:res_3}. 
In the RS regime, partial waves with $|j| > 1/2$ can also contribute due to the resonant tunneling.
These higher-$j$ partial waves can interfere with
$|j| = 1/2$ partial waves to form Fano-like resonance, see the $\rho = 6.5$ and $-4.5$ curves in Fig.~\ref{fig:lifetime}(c).
The Fano-like resonance leads to $\eta < 2$ at some electron density.
It is worth noting that $\eta$ can be smaller than unity ($\sigma_{\textrm{tr}} > \sigma$) at the Fano-like resonance,
which is unusual: it implies that the backscattering dominates the forward scattering.
Similar physics is discussed in Ref.~\onlinecite{Heinisch2013}.
However, as mentioned earlier, small randomness in size or shape of the disks unavoidable in practice would cause the ripple structure in $\eta$, including the Fano-like resonances, to be suppressed.

\section{Semiclassical ray picture}\label{sec:ray_pic}

In this Section we outline the derivation of the ray series, which is a useful
tool for investigating the $X \gg 1$ regimes, where the PWD series [Eq.~\eqref{outside_f}], although formally exact, suffers from slow convergence and does not give much physical insight.
We label the rays by integer $p$.
Figure~\ref{fig:ray_pic} shows schematically a first few rays in the series: the reflected ray $p=0$, the directly transmitted ray $p = 1$, and the ray transmitted after one internal reflection $p = 2$. The higher-order rays undergo $p - 1$ internal reflections. 
The refraction angle $\beta$ is related to the incidence angle $\alpha$ (Fig.~\ref{fig:ray_pic}) by Snell's law
\begin{equation}
\sin\beta = \dfrac{\sin\alpha}{n}\,.
\label{eq:beta}
\end{equation}
As in optics, the ray series can be derived from the PWD via an intermediate step of the so-called Debye series,  where the summation over $j$, Eq.~\eqref{outside_f},  is transformed into  an integral and evaluated by the saddle-point approximation. We applied a similar procedure to our problem, see Appendix~\ref{subsec:debye} for details. The final result has the form
\begin{subequations}
\begin{align}
f(\phi) &= \sum_{p = 0}^{\infty} f_p(\phi)
        + i\, \sqrt{\dfrac{2}{\pi k}}\, \dfrac{\sin X\phi}{\phi}\,,
\label{eqn:f_from_f_p}\\
f_p(\phi) &= e^{- \frac{i\pi}  {4}} \sum_{\alpha}
\,\left|\dfrac{d\phi_{p}}{d b}\right|^{-\frac12}
C_p(\alpha) e^{i \varphi_o + i\varphi_c},
\label{circle_ray_p}
\end{align} 
\end{subequations}
where $\phi_{p}$ is the total deflection angle of ray $p$ given by
\begin{equation}
\phi_{p} = \pi - 2\alpha - p(\pi - 2\beta)\,.
\label{eq:phi_p}
\end{equation}
Each scattering angle $\phi$ may correspond to multiple, single, or none of the rays, depending on the number of real solutions of the equation
\begin{align}
\phi = \phi_{p} + 2 \pi l\,,
\end{align}
where $l$ is some integer. 
If the dependence of $\phi_p$ on $\alpha$ is nonmonotonic,
there may be several solutions for $\alpha$ even for the same $l$,
in which case they all need to be included in the calculation of $f_p$.

Since the impact parameter (Fig.~\ref{fig:ray_pic}) is given by
\begin{equation}
b = a \sin \alpha,
\label{eqn:b}
\end{equation}
the derivative $d\phi_{p} / d b$ in Eq.~\eqref{circle_ray_p} can be written as
\begin{align}
\dfrac{d\phi_{p}}{d b} = \dfrac{1}{a \cos\alpha}
                         \dfrac{d\phi_{p}}{d\alpha}\,.
\end{align}
For now, we assume that this derivative is nonvanishing.
Later in Sec.~\ref{sec:special_angle} we explain how to modify Eq.~\eqref{circle_ray_p} if this is not the case.
The ray amplitude $C_p(\alpha)$ in Eq.~\eqref{circle_ray_p} is
expressed in terms of the plane-wave reflection and transmission coefficients of the ray at a flat interface:\cite{Katsnelson2007}
\begin{align}
&C_p(\alpha)=\left\{\begin{array}{lc}R_{\textrm{out}}&,\quad p=0\,,\\
T_{\textrm{out}}T_{\textrm{in}}R_{\textrm{in}}^{p-1}&,\quad p>0\,,\end{array}
\right.\label{eq:C}\\
&R_{\textrm{out}}=ie^{i\alpha}
\dfrac{\sin\Bigl(\dfrac{\alpha-\beta}{2}\,\Bigr)}
      {\cos\Bigl(\dfrac{\alpha+\beta}{2}\,\Bigr)}\,,\\
&T_{\textrm{out}} = 1 + R_{\textrm{out}}\,,
\label{eq:T_out}
\end{align}
and $R_{\textrm{in}}$ ($T_{\textrm{in}}$) are obtained from $R_{\textrm{out}}$ ($T_{\textrm{out}}$) by interchanging $\alpha$ and $\beta$.
The two phases appearing in the exponential in Eq.~\eqref{circle_ray_p} are
\begin{align}
\varphi_o &= -2X\cos\alpha + 2 p X' \cos\beta\,,
\label{eqn:phi_o}\\
\varphi_c &= -\dfrac{\pi}{2}\left[p - \frac12 \left(1 + \textrm{sgn}\,\dfrac{d\phi_{p}}{d b}\right)\right]\equiv -\dfrac{\pi}{2}\,N_{c}\,.
\label{eqn:phi_caustic}
\end{align}
Here $\varphi_o$ represents the phase due to the optical path length and $\varphi_c$ is the phase shift due to passing of the caustics,\cite{landau1975} which occurs $N_c$ times. 
Finally, the last term of $f(\phi)$ in Eq.~\eqref{eqn:f_from_f_p} is the usual Kirchhoff diffraction contribution.

The ray series converge much more rapidly than the PWD  because for most rays $R_{\textrm{in}}$ and $R_{\textrm{out}}$ are appreciably less than unity. (Recall that for the normal incidence the reflection vanishes exactly.)
In particular, for small $\rho$, it suffices to consider only
$p = 0$, $1$, and $2$ terms.
To the leading order in $\rho$ the solutions for $\alpha = \alpha(\phi)$ and $\beta = \beta(\phi)$ can be analytically.
Substituting these into  Eq.~\eqref{circle_ray_p},
we obtain the first three terms of the ray series as follows:
\begin{widetext}
\begin{align}
f_{0}(\phi) &=- \dfrac{\rho}{X} \sqrt{\dfrac{a}{8}}\,
 \dfrac{\cos\dfrac{\phi}{2}}{\sin^{3/2}\dfrac{\phi}{2}}
 \exp\left[-i\left(\dfrac{\pi}{4} + \dfrac{\phi}{2}
 + 2X \sin\dfrac{\phi}{2}\right)
\right]\,,
\label{eq:ray_0}\\
f_{1}(\phi) &= \sqrt{\dfrac{X}{\rho}} \sqrt{\dfrac{a}{2}}\,
 \left[1 + \left(\dfrac{X\phi}{2\rho}\right)^2\right]^{-3/4}
 \exp{\left[i\left( - \dfrac{\pi}{4}
  +\sqrt{4\rho^2 + X^2\phi^2}\,\right)\right]}\,,
\label{eq:ray_1}\\
f_{2}(\phi) &= -\dfrac{\rho}{X} \sqrt{\dfrac{a}{8}}\,
 \dfrac{\cos\dfrac{\phi}{2}}{\sin^{3/2}\dfrac{\phi}{2}}
 \exp\left[i\left(\dfrac{\pi}{4} - \dfrac{\phi}{2}
 + 2X \sin\dfrac{\phi}{2}\right)
 \right]\,.
\label{eq:ray_2}
\end{align}
\end{widetext}
These formulas will be important for the discussion of the differential cross-sections in Sec.~\ref{sec:feature_dcrs}.  

The summation of the full ray series is possible using certain approximations.
Consider the calculation of the transport cross-section.
Neglecting the ray interference and 
the diffraction term (which is important only for forward scattering),
we arrive at
\begin{equation}
\sigma_{\mathrm{tr}} = \int\limits_{-a}^{a} d b
(1 - \cos\phi)
\sum_{p} \left| A_{p} \right|^2\,,
\label{eq:ray_tcrs}
\end{equation}  
with $b$ defined by Eq.~\eqref{eqn:b} and $A_{p}$ given by
\begin{equation}
A_{p} = C_p(\alpha) e^{i\varphi_o + i\varphi_c}.
\end{equation}
Using Eqs.~\eqref{eq:phi_p}, \eqref{eq:C}--\eqref{eq:T_out}, we obtain
\begin{equation}
\sigma_{\mathrm{tr}} 
= 2 a + a\!\!\int\limits_{-\pi/2}^{\pi/2} d\alpha \cos\alpha \left[
 R^2 \cos 2\alpha
- \textrm{Re} \dfrac{e^{i(2\beta-2\alpha)}}
                    {1 + R^2 e^{2i\beta}}\right]\,,
\label{eq:pre_tcrs_ray}
\end{equation}
where $R = |R_{\textrm{out}}|=|R_{\textrm{in}}|$.
After some changes of variable Eq.~\eqref{eq:pre_tcrs_ray} can be transformed to Eqs.~\eqref{eq:tcrs_ray} and \eqref{eq:d}.
It is possible to express the correction $\varsigma(n)$ in the latter
in terms of $K(m)$, $E(m)$ and $\Pi(z, m)$, which are the
complete elliptic integrals of, respectively, the first, the second, and the third kind.
The result is
\begin{subequations}
\begin{align}
\varsigma(n) &=-\dfrac43\, \dfrac{\textrm{sgn}(n)}{(2n - 1)^2}\,
\textrm{Re}
\Bigl[
  c_{1} E\left(\dfrac{1}{n^2}\right)
+ c_{2} K\left(\dfrac{1}{n^2}\right)
\notag\\
&\mbox{} + c_{3}\, \Pi\left(\dfrac{2n-1}{n^2},\dfrac{1}{n^2}\right)
\Bigr],
\label{eqn:d_from_elliptic}\\
c_{1} &= -n(2n-1)\left(2n^3-4n^2+5n-1\right)\,,\\
c_{2} &= (n-1)(4n^4-6n^2+7n-3)\,,\\
c_{3} &= -6(n-1)^4\,.
\end{align}
\end{subequations}

Another tractable limit is $n \to \infty$, where $\beta \to 0$ so that all the odd-$p$ rays scatter into the same final direction and interfere with each other, and similar for all the even-$p$ rays.
Equation~\eqref{eq:ray_tcrs} is modified to
\begin{equation}
\sigma_{\mathrm{tr}} = \int\limits_{-a}^{a} d b
(1 - \cos\phi) \Bigl|\sum_{p} A_{p} \Bigr|^2\,,
\label{eq:ray_tcrs_infty}
\end{equation}
while the ray amplitudes are found to be 
\begin{subequations}
\begin{align}
\Bigl|\sum_{p}A_{p}\Bigr|^2 &= \Bigl|\sum_{p\textrm{ odd}}A_{p}\Bigr|^2 + \Bigl|\sum_{p \textrm{ even}}A_{p}\Bigr|^2\,,
\label{eq:tcrs_even_odd}\\
\Bigl|\sum_{p \textrm{ odd}}A_{p}\Bigr|^2
&=\dfrac{\cos^2\dfrac{\phi}{2}}{1-\sin^2\dfrac{\phi}{2}\cos^2\Phi}\,,
\label{eq:tcrs_odd}\\
\Bigl|\sum_{p \textrm{ even}}A_{p}\Bigr|^2
&=\dfrac{\sin^2\dfrac{\phi}{2}\sin^2\Phi}{1-\cos^2\dfrac{\phi}{2}\cos^2\Phi}\,.
\label{eq:tcrs_even}
\end{align}
\end{subequations}
Substituting Eqs.~\eqref{eq:tcrs_even_odd}--\eqref{eq:tcrs_even} into
Eq.~\eqref{eq:ray_tcrs_infty},
we obtain Eq.~\eqref{eq:tcrs_inf_n} by an elementary integration.

\section{Beyond the ray picture}\label{sec:sws}

In the previous sections, we have shown the benefits of the ray series for understanding the primary features of the scattering amplitudes in the large-$X$ semiclassical regime.
Here we address some interesting secondary effects that are beyond the ray picture.
Technically, the ray series is derived using the saddle-point approximation to evaluate the contour integral leading to the Debye series (Appendix~\ref{subsec:debye}).
The additional effects can be, in principle, derived by a more accurate approximation of the same contour integral.
Thus, the resonances that produce the ripples (Sec.~\ref{sec:main_results})
can be accounted for by including contributions from not only the saddle-point but also the poles in the complex $j$ plane.\cite{Nussenzveig1969b, grandy2000}
However, below we use a simpler derivation directly from the PWD.

The saddle-point approximation is also insufficient if the ray deflection angle $\phi_p$ is a nonmonotonic function of the impact parameter $b$,
so that $d \phi_p / d b$ may vanish.
In this case Eq.~\eqref{circle_ray_p} cannot be used as it
gives a divergent result.
This problem is especially apparent for refraction index in the interval $1 < n < 2$ where the divergent contribution is not overshadowed by other, non-divergent terms.
The same divergence is encountered in the theory of rainbow in optics\cite{grandy2000}
and a common remedy for it is to
replace the saddle-point approximation by a so-called uniform approximation.\cite{chester1956}
Similar issues arise for rays with incident angle $\alpha$ close to the critical angle. This regime is realized for $|n| < 1$ and is known as the near-critical scattering.\cite{grandy2000}
Finally, the saddle-point approximation becomes inaccurate if the optical phase shift $\phi_o$ [Eq.~\eqref{eqn:phi_o}] is small,
which occurs when the scattering potential is weak.
To handle this case we use an alternative approach based on perturbation theory, either the HEA or the BA. Let us now consider each of these special regimes in more detail.
 
\subsection{Resonances}\label{subsec:res}

As explained in Sec.~\ref{sec:intro},
the contribution to the cross-sections from a partial wave of a given angular momentum $j$ is proportional to $\sin^2 \delta_{j}$, where
$\delta_{j}$ is the scattering phase shift.
This quantity can be written in the form
\begin{align}
\sin^2\delta_j&=\dfrac{1}{1+\left(\dfrac{\textrm{Im}\, s_j}{\textrm{Re}\, s_j}\right)^2}\,.
\end{align}
The exact expression for $s_j$ given in Appendix~A [cf.~Eq.~\eqref{eq:S_s}] involves a combination of Bessel functions.
It can be simplified in certain limits using suitable asymptotic expansions of these functions.
Thus, for $X \gg 1$ we can use the Debye expansion to arrive, after some algebra, at
\begin{widetext}
\begin{align}
\frac{\textrm{Im}\, s_j}{\textrm{Re}\, s_j}&\simeq
\frac{\sin(\Phi_c + \Theta_+)}{\sin(\Phi_c + \Theta_-)}\,
\exp{\left(-2\sqrt{j^2-X^2} + 2j\cosh^{-1}{\frac{j}{X}}\right)},
\label{eqn:Im_s_asym}\\
\Phi_c&\equiv \sqrt{X'^2-j^2}-j\cos^{-1}{\frac{j}{X'}}-\frac{\pi}{4}\,,
\\
\Theta_\pm &= \dfrac{\pi}{2} + \tan^{-1}
\left(\dfrac{\dfrac{\sqrt{X'^2-j^2}}{X'}}
            {\dfrac{|j|\pm\sqrt{j^2-X^2}}{X} - \dfrac{|j|}{X'}}
\right)\,.
\label{eqn:Theta_pm}
\end{align}
 \end{widetext}
The resonance occurs when $\sin^2\delta_j$ attains a maximum, i.e., when the left-hand side of Eq.~\eqref{eqn:Im_s_asym} is equal to zero.
Therefore, the resonance condition is $\sin(\Phi_c + \Theta_+) = 0$, which gives Eq.~\eqref{eq:whispering} and corresponds physically to the whispering-gallery modes.

To find the resonance condition in the opposite limit of $X\ll 1$, we use a Taylor expansion of the Bessel functions that have $X$ as an argument.
We obtain
\begin{align}
\sin^2{\delta_j}\simeq\left\{1+\left[\dfrac{4^j \Gamma(1/2+j)^2}{\pi X^{2j}}\dfrac{J_{j-1/2}(X')}{J_{j+1/2}(X')}\right]^2\right\}^{-1}\,,
\label{eq:sin_delta_1}
\end{align}
where $\Gamma(x)$ is the Euler Gamma-function.
Let $Z_{j,z}$ be the $z$th zero of one of the remaining Bessel functions,
$J_{j-1/2}(Z_{j,z}) = 0$.
We define
\begin{align}
\rho_{j,z} & \equiv Z_{j,z} - X
\label{eq:sin_delta_3}
\end{align} 
and expand $J_{j-1/2}(X') = J_{j-1/2}(X + \rho)$ to the linear order in $\rho - \rho_{j,z}$: 
\begin{align}
J_{j-1/2}(X')&\simeq -J_{j+1/2}(Z_{j,z})(\rho-\rho_{j,z})\,.
\label{eq:sin_delta_2}
\end{align} 
From Eqs.~\eqref{eq:sin_delta_1}--\eqref{eq:sin_delta_3},
we obtain
\begin{align}
\sin^2{\delta_j}&=\dfrac{\gamma^2}{(\rho-\rho_{jr})^2+\gamma^2}\,,\\
\gamma&=\dfrac{\pi}{4^{j}\Gamma(j+1/2)^2}X^{2j}\,,\label{eq:gamma}
\end{align}
which leads to Eqs.~\eqref{eq:res_1}--\eqref{eq:res_3}.
It is interesting to note that
$\gamma$ vanishes if the incident electron has zero energy ($X = 0$).
Accordingly, the lifetime $1 / \gamma$ of the resonance is infinite.
Physical, it means that Dirac quasiparticles can be trapped indefinitely inside a locally doped region embedded in the otherwise undoped graphene.~\cite{Hewageegana2008,Bardarson2009, Matulis2008}

\begin{figure*}
\begin{center}
\includegraphics[scale=1]{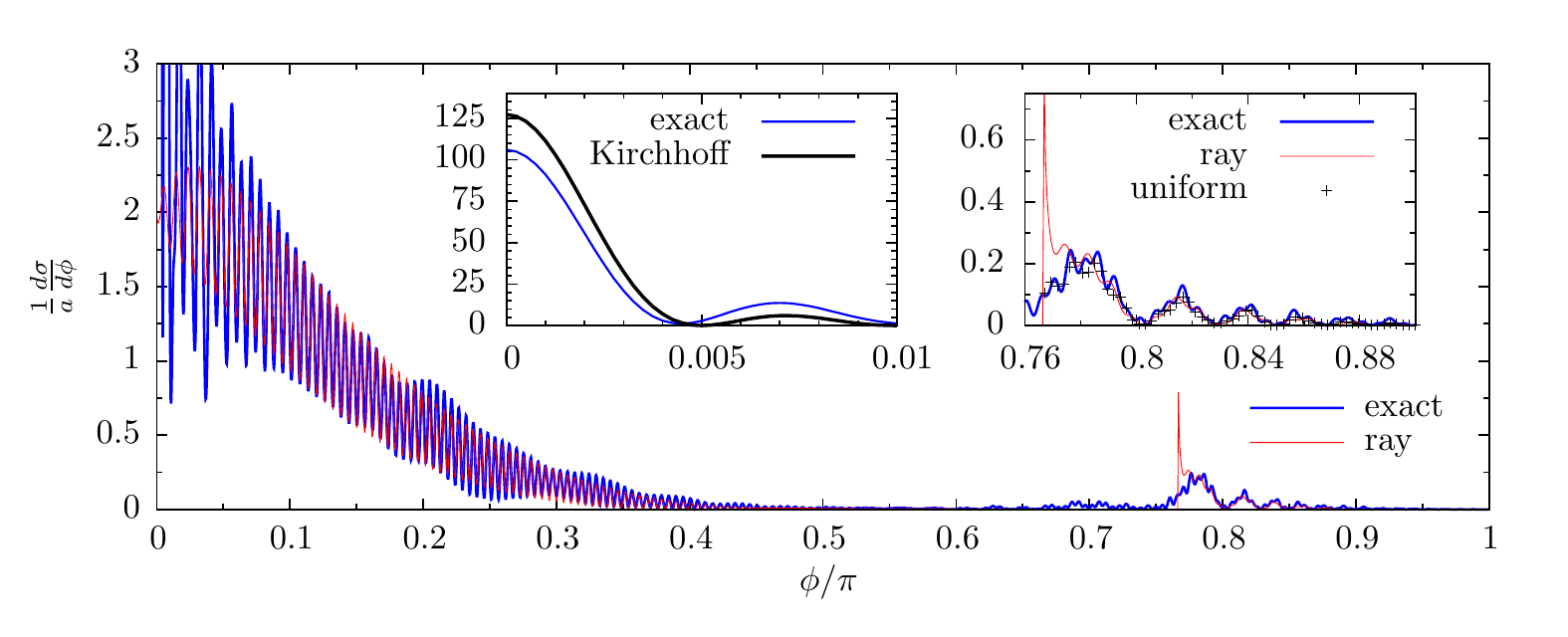}
\caption{(Color online) Differential cross-section for $X=200$ and $n=1.33$.
The main $p=2$ rainbow appears at $\phi/\pi=0.76$. 
The right inset shows  the comparison among
 the uniform approximation, the ray formula, and the PWD.
The left inset shows the comparison between the Kirchhoff diffraction  and the PWD. 
}\label{fig:rainbow}
\end{center}
\end{figure*}   

\subsection{Dirac rainbow}\label{sec:special_angle}

If $\phi_p(\alpha)$ is a nonmonotonic function of $\alpha$, there may exist $\alpha_r$ such that
\begin{align}
\left.\dfrac{d{\phi_{p}}}{d\alpha}\right|_{\alpha=\alpha_r}=0\,.
\label{eq:c_rainbow}
\end{align}
The same condition corresponds to the rainbow phenomenon in optics.
Near $\alpha = \alpha_r$, Eq.~\eqref{eq:phi_p} has more than one root.
Two of such roots, $\alpha_+$ and $\alpha_-$, coalesce at
$\alpha \to \alpha_r$.
At $\alpha=\alpha_r$, the ray formula Eq.~\eqref{circle_ray_p} diverges and cannot be used.
The divergence is cured by the
uniform approximation of the Debye series,\cite{chester1956,Berry1966,Nussenzveig1969b}
with which we obtain
\begin{align}
&f_p(\phi) = (-i)^{p } e^{i \overline\varphi} \,(-\zeta)^{1/4}
\notag\\
&\times \sum\limits_{\mu = \pm 1}
\left(\mathrm{Ai}(\zeta) + i\mu  \dfrac{\mathrm{Ai}'(\zeta)}{(-\zeta)^{1/2}}\right)
\dfrac{\sqrt{-\mu\pi}}{\sqrt{\dfrac{d\phi_{p}}{ d b}\bigg|_{\alpha_\mu}}}\,
C_p(\alpha_\mu)\,,
\label{eq:uniform_approxi_sharp}
\end{align}
where $\mathrm{Ai}(\zeta)$ is the Airy function. 
$\alpha_\pm$
are chosen such that $\varphi_o(p,\alpha+)-\varphi_o(p,\alpha-)>0$, and $\overline\varphi$ and $\zeta$ are defined by 
\begin{align}
\overline\varphi &= \dfrac{1}{2}\left[\varphi_o(p,\alpha_{+})
+
\varphi_o(p,\alpha_{-})\right]\,,\\
\zeta &=
\left\{\dfrac{3i}{4}\, \left[\varphi_o(p,\alpha_{+})
-\varphi_o(p,\alpha_{-})\right]\right\}^{2/3}.
\end{align}
An advantage of the uniform approximation is that it can be used for both $\phi < \phi_r$ and $\phi > \phi_r$ if we allow the roots $\alpha_\pm$ to be complex numbers (which are always conjugate to each other).
In doing so, the branch cut of $\zeta(\phi)$
should be chosen such that $\zeta$ is negative for real $\alpha_\pm$,
and positive when they acquire imaginary parts.

In Fig.~\ref{fig:rainbow} we compare the results of the uniform approximation,
the ray formula, and the PWD for the differential 
cross-sections computed for $n=1.33$. The rainbow condition, 
Eq.~\eqref{eq:c_rainbow}, can be satisfied for $p \geq 2$ rays.
The main $p=2$ rainbow appears at $\phi = 0.76\pi$, and the secondary $p=3$ rainbow is found at $\phi = 0.23\pi$.
As one can see from Fig.~\ref{fig:rainbow}, the ray formula strongly deviates from the exact PDW result at 
the rainbow angles.
On the other hand, at such angle the uniform approximation agrees well with 
the PWD (right inset). 
In the left inset of Fig.~\ref{fig:rainbow} we show the differential cross-section for small $\phi$ where
the ray formula also fails.
However, the differential cross-section is adequately described by the Kirchhoff diffraction formula [the last term in Eq.~\eqref{eqn:f_from_f_p}].

\subsection{Near critical scattering}

\begin{figure}
\includegraphics{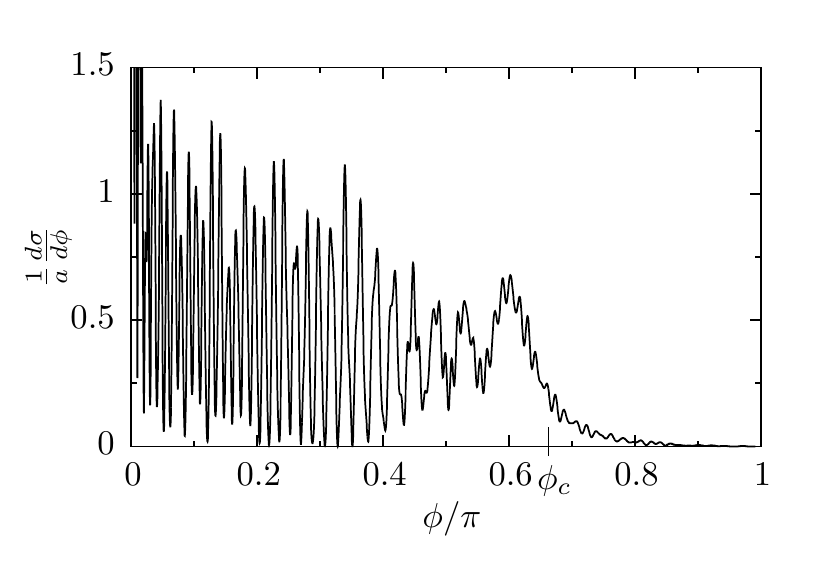}
\caption{Differential scattering cross-section for $n=0.5$ and $X = 200$ computed by the PWD. The critical angle $\phi_c = 2\pi/3$ [Eq.~\eqref{eqn:phi_c}] above which the $p = 0$ ray is totally reflected is labeled on the horizontal axis.
Near this angle the cross-section exhibits oscillations similar to those near the rainbow angle in Fig.~\ref{fig:rainbow}.
}\label{fig:critical}
\end{figure}

Another phenomenon reminiscent of rainbow is the near critical scattering,
which is realized at $|n|<1$.
At such $n$, the $p = 0$ ray is totally reflected for deflection angles larger than the critical angle
[cf.~Eq.~\eqref{eq:phi_p}]
\begin{align}
\phi_c = \pi - 2 \arcsin {n}\,.
\label{eqn:phi_c}
\end{align}
At $\phi < \phi_c$, the scattering enhanced due to the
total reflection exhibits the ``supernumerary" oscillation,
while at $\phi > \phi_c$, the scattering rapidly decays.
These Airy-function-like features are illustrated by
Fig.~\ref{fig:critical}.

\begin{figure*}
\includegraphics{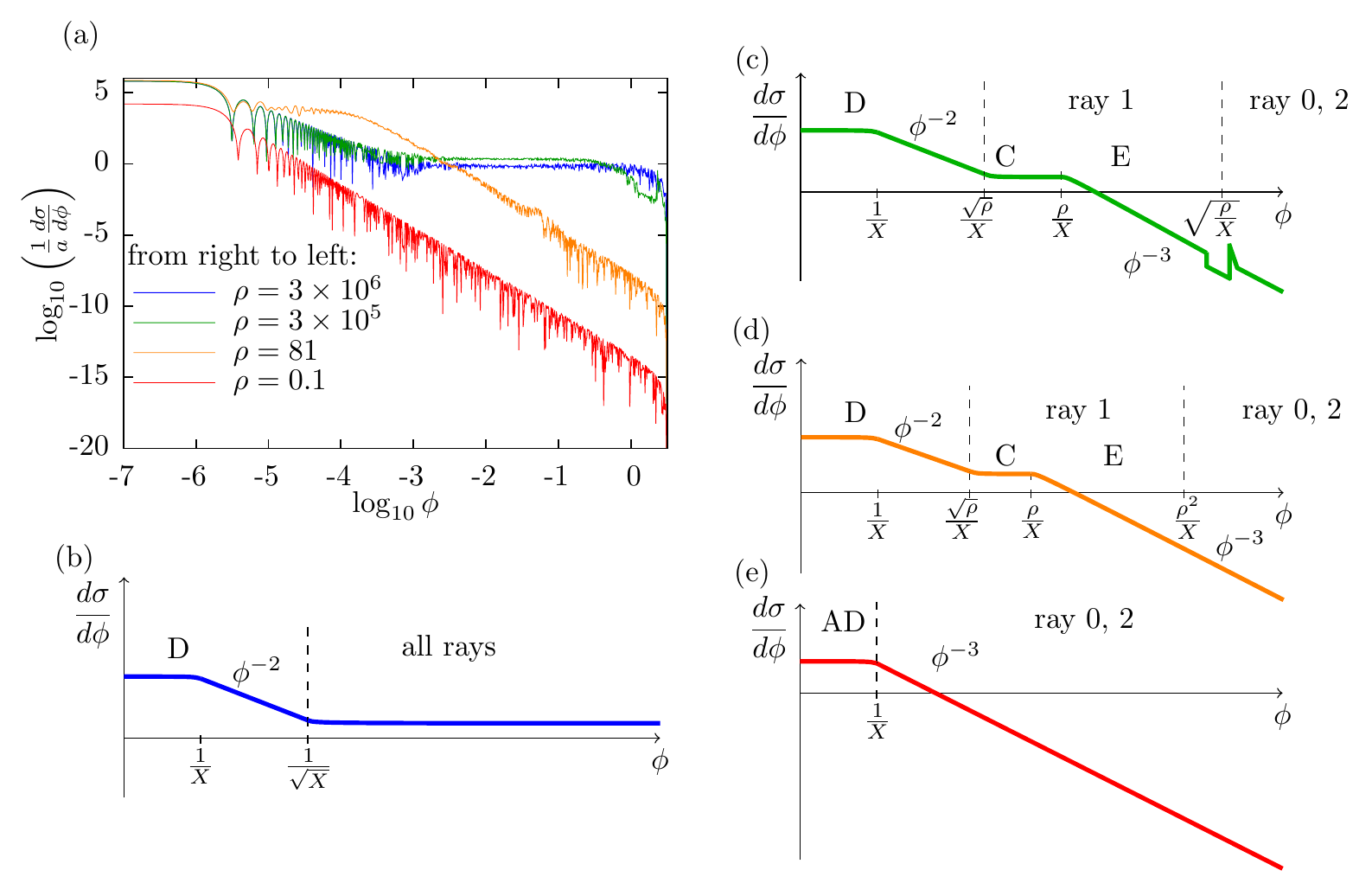}
\caption{(Color online) Differential cross-section as a function of the deflection angle. The labels on the graphs stand for D: diffraction, C: central rays ($b \ll a$), E: edge rays ($|b| \approx a$), AD: anomalous diffraction. 
(a) The exact results for $X=10^{6}$ and several $\rho$ computed using the PWD.
(b)--(e) are the schematic diagrams illustrating different regimes.
(b) SR regime. At angles less than $1 / \sqrt{X}$ the cross-section is dominated by diffraction. At larger angles, the cross-section can be computed from geometrical optics; however, accurate results require summing over many $p$ in the ray series. 
(c) WR regime. Diffraction dominates at small angles, followed by
the $p=1$ ray and then by the $p=0$ and $2$ rays.
A shallow dip followed by a peak exists in the narrow gap between the last two angular regions,
at $\phi \sim \sqrt{\rho / X}$.
The peak, which occurs on the ray-$0$,$2$ side, is the Dirac rainbow (Sec.~\ref{sec:special_angle}). 
The dip between the ray-$1$ region and the rainbow exhibits another effect beyond the ray approximation --- the Fock transition.\cite{grandy2000} 
 (d) HEA regime. Diffraction dominates at small angles followed by
the $p=1$ ray and smoothly continued by the $p=0$, $2$ rays.
(e) BA regime. Anomalous diffraction exists at angles smaller than $1 / X$. At larger angles the scattering is dominated by the $p=0$ and $2$ rays.
}\label{pic_angle_regime}
\end{figure*}

\subsection{Born approximation and high energy approximation}
\label{sec:HEA}

The development of the perturbation theory for the scattering begins with the exact Lippmann-Schwinger equation
\begin{equation}
\Psi(\mathbf{r}) = \dfrac{1}{\sqrt{2}}
 \left(\begin{array}{c}1\\1\end{array}\right)
 e^{i k x}
  + \int G_0(\mathbf{r}-\mathbf{r}')
  \widetilde{V}(\mathbf{r}')
  \Psi(\mathbf{r}')d^{2}\mathbf{r}'\,,
\label{eqn:Lippmann} 
\end{equation}
where $\widetilde{V}(r) = V(r) / (\hbar v_{F})$,
$G_0(\mathbf{r})$ is the Green's function for the 2D Dirac equation,
\begin{align}
G_0(\mathbf{r}) = -\dfrac{ik}{4}\left[\begin{array}{cc} H_{0}^{(1)}(kr)&i e^{-i\phi}H_{1}^{(1)}(kr)\\i e^{i\phi}H_{1}^{(1)}(kr)&H_{0}^{(1)}(kr)\end{array}\right],
\end{align}
and $H_j^{(1)}(x)$ is the Hankel functions of the first kind.
Following the standard route,\cite{Landau1977} we seek the solution of Eq.~\eqref{eqn:Lippmann} in the form
\begin{equation}
\Psi(\mathbf{r}) =
\dfrac{1}{\sqrt{2}}
 \left(\begin{array}{c}1\\1\end{array}\right) e^{i k x} P(\mathbf{r})\,.
\label{eqn:Psi_HEA} 
\end{equation}
The corresponding scattering amplitude is
\begin{align}
f(\phi) &= -\sqrt{\dfrac{k}{2\pi}}e^{-i\phi/2}\cos\dfrac{\phi}{2}\int \widetilde{V}(\mathbf{r}) P(\mathbf{r}) e^{-i\mathbf{q}\cdot\mathbf{r}}d^{2}\mathbf{r}\,,
\label{eq:HEA}\\
\mathbf{q} &= k(\cos\phi - 1, \sin \phi)\,.
\label{eqn:q}
\end{align}
One can show that for small angles and weak enough scattering potential, $\phi \ll \rho / X \ll 1$, function $P$ is approximately equal to
\begin{equation}
P(x, y) = \exp{\left(-i\int_{-\infty}^{x}
\widetilde{V}(x',y)dx'\right)}\,.
\label{eqn:P} 
\end{equation}
This approximation~\cite{Landau1977} is known as the HEA.
(A similar method called ``paraxial approximation" was applied to graphene in Ref.~\onlinecite{Matulis2011}.)
Since $\phi$ is always considered to be small within the HEA, we can achieve further simplification by writing the change in momentum [Eq.~\eqref{eqn:q}] as
\begin{equation}
\mathbf{q} \simeq k \phi \hat{y}\,.
\label{eq:q_perp}
\end{equation} 
Substituting this into Eq.~\eqref{eq:HEA} and integrating over $x$,
we obtain
\begin{align}
f(\phi) &=-i\sqrt{\dfrac{k}{2\pi}}e^{-i\phi/2}\cos\dfrac{\phi}{2}
\notag\\
&\times \int_{-\infty}^{\infty} \left[ P(\infty, y) - 1\right]
 e^{-i k \phi y} d{y}\,.
\label{eq:HEA_final}
\end{align}
Furthermore, under the condition $\int \widetilde{V} d r \ll 1$,
the quantity $P(\infty, y) - 1$ can be expanded to the first order in $V$,
which leads to the first Born approximation (BA).
Within the latter, the formula for the scattering amplitude
can be simplified by integrating over the polar angle:
\begin{equation}
f(\phi) = -\sqrt{\dfrac{k}{2\pi}}e^{-i\phi/2}
\cos\dfrac{\phi}{2}
\int_{0}^{\infty}2\pi \widetilde{V} (r) J_{0}(q r)r d r\,.
\label{eq:BA}
\end{equation} 
Although the HEA is valid only for small $\phi$, the BA result holds for any $\phi$ as long as the potential $V$ is weak, as specified above,
and we use Eq.~\eqref{eqn:q} for $\mathbf{q}$, i.e., $q = |\mathbf{q}| = 2 k \sin (\phi / 2)$.
For the step potential [Eq.~\eqref{eq:potential}], the calculations according to Eqs.~\eqref{eq:HEA_final}, \eqref{eq:BA}, and
\eqref{eq:op_theorem} yield the formulas for the total cross-sections presented in Sec.~\ref{sec:main_results}.
Computing the transport cross-sections within the HEA and BA [Eqs.~\eqref{eq:tcrs_HEA} and \eqref{eq:tcrs_BA}]
requires deriving the differential cross-sections first.
These quantities are discussed in the next Section. 

\section{Differential cross-sections in each regime}
\label{sec:feature_dcrs}

In principle, numerical evaluation of the PWD series is not difficult even for large $X$ as long as one keeps enough terms in the summation and uses reliable library subroutines for computing the requisite Bessel functions.
The results of these calculations, which can be considered numerically exact, for $X = 10^6$ are presented in Fig.~\ref{pic_angle_regime}(a).
While it is too high to be practical for graphene,
choosing such a large $X$ enables us to show more clearly the qualitative trends displayed by the differential cross-section $d\sigma / d\phi$ as a function of $\phi$ for four different fixed $\rho$. 
These trends are schematically illustrated in Figs.~\ref{pic_angle_regime}(b)--(e).
In the rest of this Section we discuss how these trends can be understood based on the analytical approximations we developed above.

We begin with the SR regime where the ray series is accurate,
and so Eqs.~\eqref{eqn:f_from_f_p} and \eqref{circle_ray_p} can be used.
At small angles, 
the last (diffraction) term in Eq.~\eqref{eqn:f_from_f_p} is the dominant one.
It goes to a constant at $\phi = 0$ and decays as $d\sigma / d\phi \sim \phi^{-2}$ at $\phi \gg 1/X$.
A crude sketch illustrating this behavior is
shown in Fig.~\ref{pic_angle_regime}(b). It consists of a plateau at $\phi<1/X$ and a straight line at $\phi > 1/X$.
We label it by ``D'' (for ``diffraction'').
The ray series dominates the diffraction at $\phi>1/\sqrt{X}$, 
as shown schematically in Fig.~\ref{pic_angle_regime}(b).
Since the scattering potential is strong, there is no particular restriction on the characteristic deflection angle of the rays. For any $p$, it can be as small as zero or as large as the maximum possible angle $\pi$.
Therefore, the differential cross-section due to the ray contribution is shown to be flat in Fig.~\ref{pic_angle_regime}(b), meaning it does not vary much on the logarithmic scale.
This is indeed the behavior exhibited by the $\rho = 3 \times 10^6$ curve in Fig.~\ref{pic_angle_regime}(a), which corresponds to the
largest refraction index curve ($n = 4$) in that Figure.

In the WR regime,
the ``D'' feature is also present, see Fig.~\ref{pic_angle_regime}(c).
The ray contribution exceeds that of diffraction at $\phi > \sqrt{\rho}/X$.
From Eqs.~\eqref{eq:ray_0}--\eqref{eq:ray_2}, one can show that
the $p=1$ ray dominates at $\sqrt{\rho}/X<\phi<\sqrt{\rho/X}$ and the $p =0$ and $p=2$ rays take over at $\phi>\sqrt{\rho/X}$.
The contribution of the $p=1$ ray to the cross-section decays at $\phi > \rho/X$ as $d\sigma / d\phi \sim \phi^{-3}$.
This occurs because at such $\phi$ the $p = 1$ rays graze along the edge of the disk, $b \approx a$; hence, this domain is labeled ``E'' in Fig.~\ref{pic_angle_regime}(c).
Note that a small dip in $d\sigma / d\phi$ exists at $\phi\sim\sqrt{\rho/X}$ that separates angular intervals dominated by $p=1$ ray and the $p=0$, $2$ rays.
Inside this dip no classical solutions exist for either $p = 1$ or $2$ ray,
while the contribution of $p = 0$ ray is already small.
However, on the high-$\phi$ side of the dip there are two such solutions for $p = 2$.
The interference between them gives rise to the Dirac rainbow, which was discussed in Sec.~\ref{sec:special_angle}.

In the HEA and BA regimes, Figs.~\ref{pic_angle_regime}(d) and \ref{pic_angle_regime}(e), respectively, the formulas from Sec.~\ref{sec:HEA} apply.
In the HEA regime, we use Eqs.~\eqref{eq:potential} and \eqref{eq:HEA_final} to obtain
\begin{align}
f(\phi) &= i\, \sqrt{\dfrac{2}{\pi k}}\, \dfrac{\sin X\phi}{\phi}
+ I(\phi)\,,
\label{eq:f_HEA}\\
I(\phi) &= -i\sqrt{\dfrac{k}{2\pi}} \int\limits_{-1}^{1}dy\,
e^{-i{X \phi y}} \exp{\left( 2i \rho\sqrt{1 - y^2}\right)}\,.
\label{eq:I}
\end{align}
The first term of Eq.~\eqref{eq:f_HEA} is 
the same as the diffraction term in Eq.~\eqref{eqn:f_from_f_p}.
The second term $I(\phi)$ admits analytical approximations in some limits.
At $\phi\ll \rho^2/X$, it can be calculated by the saddle-point approximation.
The result is identical to $f_1(\phi)$ given by Eq.~\eqref{eq:ray_1}.
At $\rho^2/X \ll \phi \ll 1$, the leading-order analytical approximation to $I(\phi)$ can be calculated by deforming the contour to a rectilinear path $(-1, -1 -  i\nu  \infty, 1 - i\nu \infty, 1)$, with $\nu = \textrm{sgn}\, \rho$.
The result is
\begin{align}
I(\phi) &\simeq -\dfrac{2\rho}{X}\, \sqrt{\dfrac{a}{\phi^3}}\,
\cos\left(X\phi + \dfrac{\pi}{4}\right)\,,
\label{eq:ray_02}
\end{align}
which is equal to the sum of $f_0(\phi)$ and $f_2(\phi)$, Eqs.~\eqref{eq:ray_0} and \eqref{eq:ray_2}.
In the intermediate region, $\phi \sim \rho^2/X$,
the two approximations match by the order of magnitude but none of them is quantitatively accurate. If desired, the integral $I(\phi)$ can be calculated numerically.
The result would then provide a smooth connection between the $p = 1$ and $p = 0$, $2$ ray formulas. 
Unlike in the WR regime, the differential cross-section has neither a dip nor a peak (rainbow) in this regime. Such geometrical optics features are smeared out in the HEA regime. 
Apart from this, it is remarkable that the HEA, which is typically considered a quantum theory, can be reproduced using the semiclassical ray picture.
It is not very difficult to show that the crossover between the HEA and the WR occurs at $X \sim |\rho|^3$, see Fig.~\ref{fig:regime}.

The differential cross-section of the HEA regime is shown schematically in Fig.~\ref{pic_angle_regime}(d). 
The $p=1$ ray dominates over diffraction
at $\phi > \sqrt{\rho}/X$.
Its contribution to the cross-section behaves as
 $\phi^{-3}$ at $\phi > \rho / X$.
The same trend is smoothly continued by
the $p = 0$, $2$ rays, which dominate at $\phi > \rho^2 / X$.

Lastly, in the BA regime, we use
Eqs.~\eqref{eq:potential} and \eqref{eq:BA} to obtain
\begin{equation}
f(\phi) = \sqrt{2\pi X a}\, \rho\,
e^{-i\phi/2} \cos\dfrac{\phi}{2}
\dfrac{J_{1}\left(2X\sin\dfrac{\phi}{2}\right)}{2X\sin\dfrac{\phi}{2}}\,,
\label{eq:f_BA}
\end{equation}
which entails Eq.~\eqref{df_BA}.
The differential cross-section in the BA regime is shown in Fig.~\ref{pic_angle_regime}(e).
As mentioned in Sec.~\ref{sec:main_results},
the small-angle scattering in the BA regime is described by the anomalous diffraction, which originates from the 
destructive interference of the $p=1$ rays with the usual diffraction.
Consequently, the maximum value $\sigma'(0) = (\pi / 2) a \rho^2 X$ of the differential cross-section is much smaller
than the Kirchhoff result $\sigma'(0) = (2 / \pi) a X$
[see the first term in Eq.~\eqref{eq:f_HEA}]. At ${1}/{X} \ll \phi \ll 1$, Eq.~\eqref{eq:f_BA} agrees with 
Eq.~\eqref{eq:ray_02} because the BA and the HEA are both valid at such angles, predicting the $d\sigma / d\phi \propto \phi^{-3}$ decay.

In summary, scattering of quasiparticles by large disks, $X \gg 1$,
can be described by the ray series at all but very small deflection angles $\phi$. 
At such small angles, there is a competition between the rays and diffraction.
Diffraction dominates for the strong enough potential, $\rho \gg 1$. In the opposite case, the $p=1$ rays nearly cancel the diffraction, making it ``anomalous.''
As one can see from Fig.~\ref{pic_angle_regime}(a),
the exact results for the differential cross-section at sufficiently large 
$X = 10^6$ agree very well with our schematic diagrams for all $\rho$ pictured therein. 

Based on the above results for the differential cross-sections,
derivation of the transport cross-sections within the HEA and BA [Eqs.~\eqref{eq:tcrs_HEA} and \eqref{eq:tcrs_BA}] is straightforward,
and so we will not elaborate on it.

\section{Near-field scattering}
\label{sec:nfs}

\begin{figure}
\includegraphics{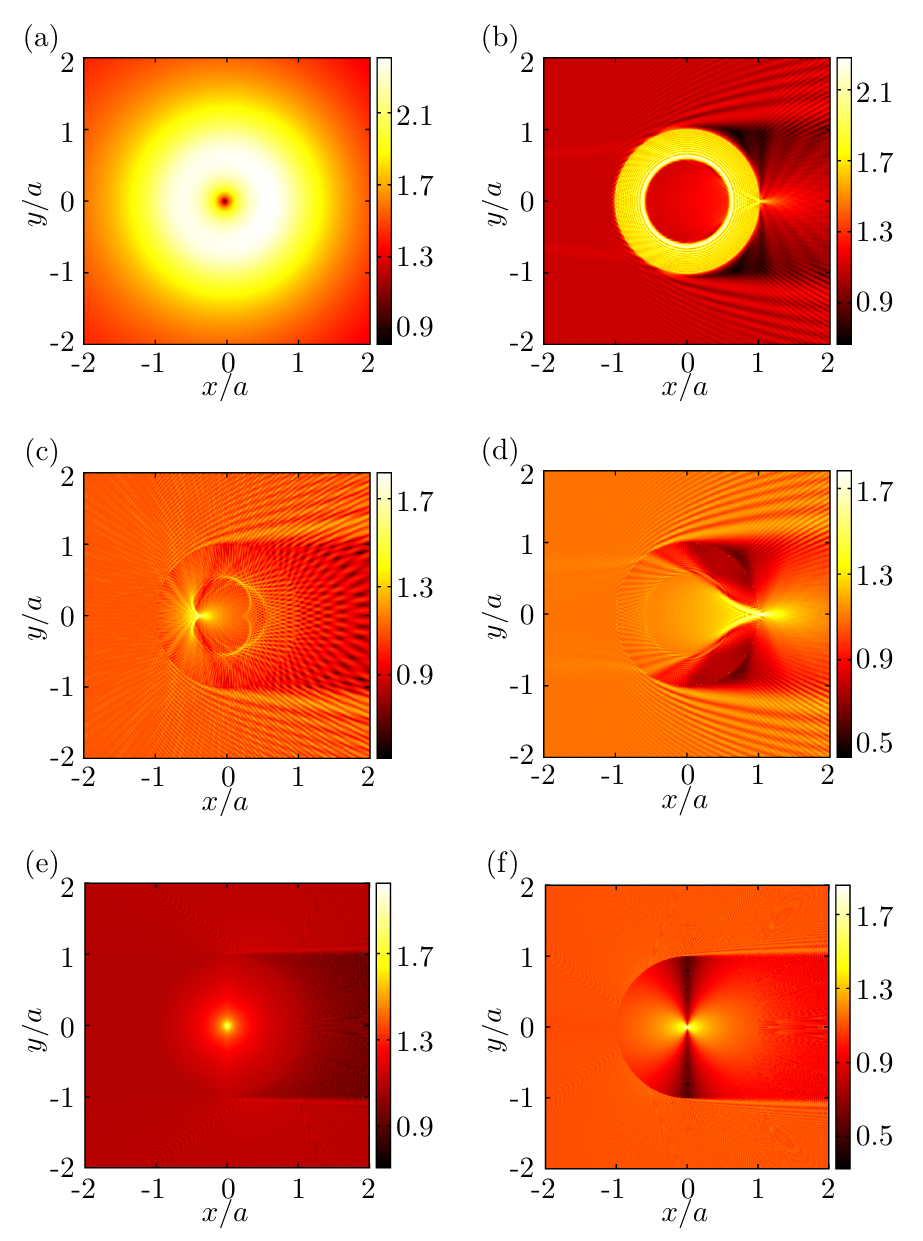}
\caption{(Color online) Near-field features of the scattering wavefunction $\Psi(x, y)$ for several representative choices of $X$ and $\rho$.
In order to avoid drastic contrast variation in these false color diagrams,
we plot $|\Psi|^{1/8}$ rather than $|\Psi|^2$.
(a) A low angular momentum resonance occurring at $X=0.3$, $\rho=3.363$. 
The partial waves with $j=\pm{3}/{2}$ are resonant.
(b) A high angular momentum resonance (``whispering gallery mode'') at $X=100$, $\rho=81.7762$. The partial waves with $j=\pm{165}/{2}$ are resonant.
(c) An example of negative refraction in the VR regime, $X=100$, $\rho=-281$ ($n=-1.81$).
(d) An example of caustic and lensing in the SR regime, $X=100$, $\rho=81$ ($n=1.81$).
(e), (f) FP regime $n \gg 1$ on and off the resonance, $\{X' / \pi\}=0$ and $0.25$, respectively (cf.~Fig.~\ref{data_inf_n}).
}
\label{fig:near}
\end{figure}

\begin{figure}
\includegraphics[width=8.8cm]{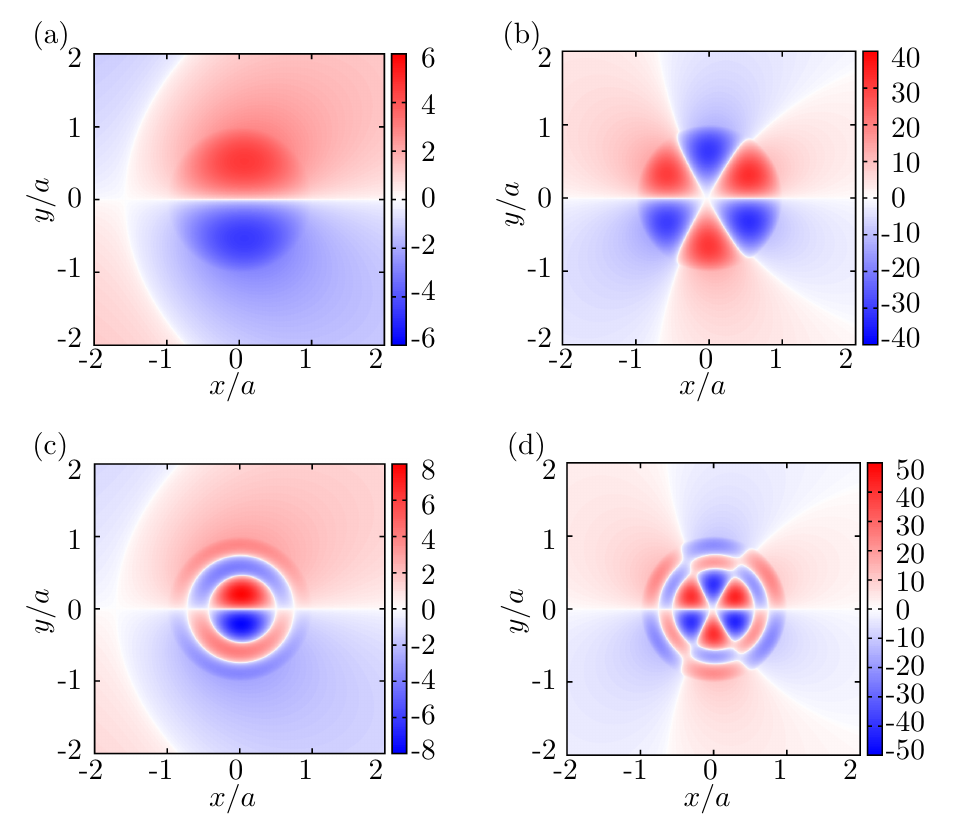}
\caption{(Color online) Local $z$-component spin density
$\langle\sigma_z\rangle \equiv \langle \Psi| \sigma_z | \Psi \rangle$
for the four resonances at $X = 0.3$ shown in Fig.~\ref{fig:data_angle_dcrs}. 
We plot $\textrm{sgn}\langle\sigma_z\rangle |\langle\sigma_z\rangle|^{1/2}$ rather than $\langle\sigma_z\rangle$ to avoid too much contrast in these false color diagrams.
(a) $\rho = 1.739$, the $j = {1}/{2}$, $z = 1$ resonance. 
(b) $\rho = 3.363$, the $j = {3}/{2}$, $z = 1$ resonance. 
(c) $\rho = 4.883$, the $j = {1}/{2}$, $z = 2$ resonance. 
(d) $\rho = 6.555$, the $j = {3}/{2}$, $z = 2$ resonance. 
}
\label{fig:nf_spin}
\end{figure}

The previous Sections have been devoted to quantitative and qualitative discussion of far-field scattering amplitudes.
In this Section,
we turn our attention to the behavior of the electron wavefunction $\Psi(x, y)$ near the scatterer.
As previously, we assume that electrons propagate ballistically before and after they collide with the disk.
In such an idealized system, the incident plane wave can be created by
injecting a small electric current in a particular direction.
One quantity we will discuss is $|\Psi(x,y)|^2$,
which determines the current-induced change in the local charge density (CCLCD).
Additionally, having in mind Dirac fermions on a surface of a topological insulator, we will consider $\langle \sigma_z \rangle \equiv \langle \Psi| \sigma_z | \Psi\rangle$,
which determines the current-induced change of the $z$-component of the local \textit{spin} density (CCLSD).
In graphene, the same expectation value defines pseudospin rather than spin density (and so it may be more difficult to probe experimentally).
Our goal is to see how the qualitative features of the CCLCD and the CCLSD can be understood in terms of the concepts introduced in the previous Sections, in particular,
partial wave resonances and semiclassical ray trajectories.

Figure~\ref{fig:near} shows a suitably normalized CCLCD for six
representative choices of $X$ and $\rho$.
The false color scale in this Figure is effectively nonlinear
because we plot $|\Psi|^{1/8}$ instead of $|\Psi|^2$ to mitigate sharp contrast variations.
The top left panel of Figure~\ref{fig:near} pertains to the smallest-$\rho$ $|j| = 3/2$ resonance in the RS regime.
The resonance is seen to generate a dramatic enhancement of the CCLCD at the scatterer.
It may be surprising that this enhancement is axisymmetric.
This defies
naive expectations that interference of $\pm j_r$ partial waves should produce an oscillatory angular dependence of the CCLCD.
In fact, the lack of angular oscillations is peculiar to massless Dirac fermions.
The states of angular momenta $\pm j_r$
are characterized by mutually orthogonal spinors;
therefore, these states do not interfere with one another yielding a nearly axisymmetric CCLCD.

Resonances can also occur in the WR and SR regimes for the partial waves
of high angular momenta $|j| \gg 1$, cf.~Eq.~\eqref{eq:whispering}.
Such partial waves
are analogous to the ``whispering gallery'' modes in optics.
They produce ring-shaped CCLCD enhancement
shown in Fig.~\ref{fig:near}(b).
(Note that within the semiclassical picture the resonant states correspond to particles trapped inside the disk and orbiting in either direction around its center.)
In Figs.~\ref{fig:near}(c) and \ref{fig:near}(d),
we present examples of negative\cite{Cheianov2007} and positive refraction, respectively.
Figure~\ref{fig:near}(c) depicts the CCLCD for $n = -1.81$, which is in the VR regime.
The most notable features are the internal caustics that can be explained using the ray picture.\cite{Cserti2007}
Figure~\ref{fig:near}(d) shows CCLCD for $n = 1.81$.
Here the refraction is positive and
instead of the internal caustics, the rays exhibit  
focusing outside the disk.
The narrowness of the whispering-gallery resonance can be appreciated by
comparing Figs.~\ref{fig:near}(b) and \ref{fig:near}(d).
Although the refractive indices in the two cases differ by little more than $1\%$,
their CCLCD look dramatically different.
While in Fig.~\ref{fig:near}(d), the CCLCD is dominated by the focal point of the rays, 
in Fig.~\ref{fig:near}(b) it is almost completely overshadowed by the resonant partial waves.
Thus, for the CCLCD 
a single pair of resonant partial waves can be more prominent than
all other waves combined.
This is in contrast to the far-field scattering quantities at $X \gg 1$, for which
such a resonance typically gives only a small correction to the ray formula result [Eqs.~\eqref{eq:4a} and \eqref{eq:res_2}].

The last two panels in Fig.~\ref{fig:near} illustrate the role of ray interference in the FP regime ($n \gg 1$), where the far-field cross-section exhibits periodic oscillations, see Fig.~\ref{data_inf_n} in Sec.~\ref{sec:main_results}.
We see in Figs.~\ref{fig:near}(e) and \ref{fig:near}(f) that the ray interference also strongly influences the CCLCD, causing marked change in the CCLCD intensity along the vertical diameter of the disk on and off the resonance. 

Let us now turn to the features of the local spin density.
We have found that the CCLSD maps in the WR and SR regimes show qualitatively the same caustics and focal spots as the maps of the CCLCD.
Therefore, these CCLSD maps do not seem to give much additional information and are not presented here.
However, striking differences between the CCLCD and CCLSD appear in the RS regime.
Figure~\ref{fig:nf_spin} shows the CCLSD at the positions
of the four resonances seen in Fig.~\ref{fig:data_angle_dcrs}.
Unlike the CCLCD maps, the CCLSD shows strong angular variations.
This can be seen from the comparison of Figs.~\ref{fig:near}(a) and
\ref{fig:nf_spin}(b), which have the same $\rho$ and $X$.
In order to avoid too drastic contrast variations in the false color,
we again use nonlinear scaling and plot
$\textrm{sgn}\langle\sigma_z\rangle|\langle\sigma_z\rangle|^{1/2}$ instead of $\langle\sigma_z\rangle$ in Fig.~\ref{fig:nf_spin}.
The oscillations of the CCLSD are enabled by the discussed above interference of the $\pm j_r$ waves owing to the nonzero term $\langle \Psi_{-j_r}|\sigma_z|\Psi_{j_r}\rangle$.
This interference causes the CCLSD to change its sign $2 |j_r|$ times in the angular direction
and $z$ times in the radial direction.

One may wonder how the predicted CCLCD or CCLSD can be measured in experiments.
We think it may be possible with modern scanned-probe techniques.
However, this task would require probing a current-carrying system with a nanoscale spatial resolution.
One somewhat indirect method is to utilize the scanned gate microscopy (SGM),
in which the change of the conductance of the system is measured in response to a local gating by the scanned tip.
With further analysis, this type of measurement can in principle reveal the CCLCD.\cite{Eriksson1996, Topinka2000, Topinka2001, Berezovsky2010, Sellier2011,Mre2014}
A more direct method is the scanned tunneling potentiometry (STP)~\cite{Muralt1986} recently implemented to study current-carrying graphene.\cite{Wang2013}
By incorporating magnetized scanned probes into the SGM and STP,
it may also become possible to study the predicted patterns of the local spin density
on a surface of a topological insulator.

\section{Discussion and conclusions}
\label{sec:conclusion}

Several remarks are in order before we conclude.
First, the transport properties of graphene have provided a major motivation for this study. Since there have already been numerous previous theoretical investigations 
of this subject,
it may be worthwhile to draw attention to the points where we find qualitatively different results.
Recall that the two most common models of scatterers in graphene are
random (uncorrelated) charged impurities and random short-range defects.
For the former, the theory predicts the linear dependence of the conductivity
on the electron density $n_e$, for the latter the conductivity is roughly density-independent.\cite{Peres2010ctp, DasSarma2011eti}
Introducing some degree of correlations among the impurity positions into these basic models
modifies the conductivity dependence quantitatively~\cite{Li2011} but preserves this main dichotomy. 
In contrast, in our model the conductance has an approximately $\sqrt{n_e}$ dependence 
[Eq.~\eqref{eq:conductance}] if the potential barrier is strong enough,
$\rho > 1$. 
Let us give a specific example.
Suppose the potential scatterers of our model are formed by aggregation of charged impurities with average density $10^{13}\,\mathrm{cm}^{-2}$
into circular clusters inside of which the distances between the impurities
is about $1\,\mathrm{nm}$.
From the conservation of the total impurity number one concludes that in this model there is an inverse relation between
the density $n_c$ of the clusters and their radius $a$:
\begin{equation}
 n_c \sim 0.03 a^{-2}\,.
\label{eqn:n_c_from_a}
\end{equation}
Substituting this formula into Eq.~\eqref{eq:conductance},
we obtain an estimate of the conductivity
\begin{equation}
G \sim 100 \frac{e^2}{h}\, \sqrt{n_e} a\,,
\label{eqn:g_100}
\end{equation}
which exhibits the $\sqrt{n_e}$-behavior.
Since $G$ is proportional to $a$, formation of clusters greatly increases the conductivity\cite{Katsnelson2009} in comparison to the case of random isolated impurities.
This conclusion is further strengthened by the large numerical factor in Eq.~\eqref{eqn:g_100}.

The lifetime ratio $\eta$ [Eq.~\eqref{eq:eta_def}] is another fundamental parameter characterizing the transport properties of the system.
Most of previous calculations of $\eta$ have been limited to the perturbation theory (the Born approximation), which predicts $\eta \geq 2$ for the charged and $1 \leq \eta \leq 2$ for the short-ranged
defects.
As we point out in Sec.~\ref{sec:main_results} and
illustrate in Fig.~\ref{fig:lifetime}(c),
$\eta$ can be less than $2$ and even less than $1$ for small-radius scatterers because of the resonant tunneling, a non-perturbative effect.
Conversely, for large-radius scatterers, $\eta$ can be very large, see Fig.~\ref{fig:lifetime}(a).
Note that the small lifetime ratio $\eta < 2$ observed in some experiments\cite{Hong2009, Tiras2013} while
large $\eta \sim 6$ is found in some others.~\cite{Gorbachev2009}

Next, we wish to address the validity of our step-like model of the potential barrier.
If the potential is indeed created by a cluster of charged impurities,
this model is oversimplified.
The actual potential has no discontinuity.
Instead, it sharply but continuously drops over a distance of the order of the screening length, which is usually comparable to the Fermi wavelength.\cite{Castro2009}
For such a smooth boundary,
the reflection and transmission coefficients that enter
the ray formula Eq.~\eqref{circle_ray_p} are modified,
e.g., the reflection coefficient is enhanced.\cite{Cheianov2006, Zhang2008}
Therefore, the transport and total cross-sections should be greater than what we calculated for a sharp boundary.
The correction is relatively small
if both $X$ and $X'$ are large, so that the radius of the cluster is much larger than the Fermi wavelength on both sides of the boundary.
However, if either exterior or interior of the cluster is doped weakly, a more accurate calculation will be necessary.
Another omission of the step-like model
is the long-range tail of the screened Coulomb potential induced by the cluster.
For high Fermi energy the screening is strong and the effect of such a tail can be treated perturbatively.
A weak long-range potential tail would cause additional small-angle scattering, which should make only a small correction to the transport cross-section.
The effect on the total cross-section could be more substantial.
Within the HEA, the contribution of the long-range potential tail to the total cross-section is given by \cite{Landau1977}
\begin{align}
\Delta\sigma &\simeq \dfrac{8 a}{X} \int\limits_{0}^{\infty}
[\sin^2 (\delta_j + \Delta\delta_j) - \sin^2 \delta_j]\, d j\,,
\label{eqn:sigma_tail}\\
\Delta\delta_j &\simeq -\dfrac{1}{\hbar v_F} \int\limits_{\mathrm{max}(a, b)}^{\infty}
\dfrac{{V}(r) r d r}{\sqrt{r^2 - b^2}}\,,
\quad b = \dfrac{j a}{X}\,.
\label{eqn:delta_tail}
\end{align} 
It is well known that an external charge screened by a gas of 2D electrons
(either massive or massless) produces the potential that behaves as $V(r) \propto r^{-3}$ 
at distances much larger than the screening length.
It is then easy to see that the integral in Eq.~\eqref{eqn:sigma_tail} converges.
Thus, there should be a range of parameters where neglecting $\Delta\sigma$ is legitimate.
However, for massless Dirac fermions, large enough total charge of the cluster,
and low enough Fermi energy, the screened potential would exhibit a slower decay $V(r) \propto r^{-3/2}$ over a range of intermediate distances.\cite{Fogler2007}
According to Eq.~\eqref{eqn:sigma_tail}, this may yield logarithmic corrections to the total cross-section.
This interesting problem warrants further study.

In conclusion, we studied in some depth scattering of massless Dirac fermions by a step-like circular potential.
We unified many possible scattering regimes into a single global diagram (Fig.~\ref{fig:regime}) and presented analytical and numerical results for the scattering amplitude in each of the individual regimes.
We verified that the semiclassical ray formalism accounts for most of the scattering properties in the large-$X$ regimes and at the same time provides an intuitive physical description of both the far-field and near-field scattering.
We showed that the ray picture applies even for weak scattering potentials,
which is the case where the semiclassical method is usually eschewed in favor of quantum scattering theory.
We also discussed phenomena beyond the ray picture,
such as the regularization of the divergence of the scattering amplitude at the rainbow angle and the quantization of the whispering-gallery resonances.
While the perfect axial symmetry and the step-like discontinuity of the potential barrier that enabled us to make progress in terms of analytical theory are not fully realistic,
some of our nonperturbative semiclassical techniques can be
extended to barriers of more general shapes and profiles.
It may also be interesting to apply our techniques to
bilayer graphene and graphene with an externally-induced mass.\cite{Matulis2008, Masir2011b}
We expect the types of the scattering regimes to be the same.
However, because of the energy-dependent Fermi velocity,
different pseudospin structure, and/or
suppression of the Klein tunneling
by the mass gap in these systems the positions of the regime boundaries and
the angular dependence of the differential cross-sections should be different from those for monolayer graphene.
Finally, we hope that our results may stimulate future transport and scanned-probe experiments with graphene and topological insulators.

\begin{acknowledgments}

This work is supported by Study Abroad Fellowship of MOE Taiwan (R.O.C.),
by ONR under Grant No.~N0014-13-0464, by University of California Office of the President.

\end{acknowledgments} 

\appendix
\section{Partial wave decompostion}\label{subsec:pwd}

In order to make the paper self-contained, in this Appendix we review the partial wave series.
Using the notations of Ref.~\onlinecite{Cserti2007}
we denote by $h_{j}^{(2)}$ and $h_{j}^{(1)}$, respectively, the incoming and the outgoing waves of angular momentum $j$.
At $r > a$, where $V(r)=0$, such waves are given by
\begin{align}
h_{j}^{(d)}(r,\phi)=\left(\begin{array}{c}
                  H_{j-1/2}^{(d)}(k r)e^{-i\phi/2}\\
                  i H_{j+1/2}^{(d)}(k r)e^{i\phi/2}
                  \end{array}\right)e^{i j \phi}.
\label{eq:PWD_hankel}
\end{align}
The angular momenta $j$ are half integers, so that $h_{j}^{(d)}$ are single-valued.
At $0 \leq r < a$, the solutions that are well-behaved at the origin are
\begin{align}
\chi_{j}(r,\phi)=\left(\begin{array}{c}
                  J_{j-1/2}(k' r)e^{-i\phi/2}\,\\
                  i J_{j+1/2}(k' r)e^{i\phi/2}
                  \end{array}\right)e^{ij\phi}
\label{eq:PWD_inside}
\end{align}
with $k' = {(E - V_0)}/{(\hbar v_F)} = X' / a$.
Equation~\eqref{eq:PWD_inside} can be used for both  positive and
negative $k'$. In the latter case,  a $n$-$p$ junction forms at the boundary of the scatterer. 

In the partial wave method, the scattering wavefunction $\Psi$ is expanded as follows.
At $r > a$, it is given by
\begin{align}
\Psi = \dfrac{1}{\sqrt{2}}
 \left(
 \begin{array}{c}1\\1\end{array}
 \right) e^{i k x} + \dfrac{e^{-\frac{i\pi}{4}}}{2\sqrt{2}}
 \sum_{j} i^{j} \left(e^{2 i \delta_{j}} - 1\right) h_{j}^{(1)}\,,
\label{outside}
\end{align} 
where the coefficient in front of the sum is chosen to match the coefficient in a similar expansion of the incident plane wave (the first term).
At $r < a$, we have
\begin{align}
\Psi = \dfrac{e^{-\frac{i\pi}{4}}}{2\sqrt{2}}
\sum_{j} i^{(j-1/2)} B_{j}\chi_{j}.
\label{inside}
\end{align}
By imposing the continuity of the wavefunction at $r=a$,
it is straightforward to find
\begin{align}
e^{2i\delta_{j}} & =-\dfrac{s_{j}^{*}}{s_{j}}\,,
\label{eq:S_s}\\
s_{j} &= H_{j+1/2}^{(1)}(X) J_{j-1/2}(X')- H_{j-1/2}^{(1)}(X) J_{j+1/2}(X')\,,
\label{eq:s_H}\\
B_{j} &= \dfrac{H_{j+1/2}^{(1)}(X) H_{j-1/2}^{(2)}(X)
              - H_{j-1/2}^{(1)}(X)  H_{j+1/2}^{(2)}(X)}
               {s_{j}}\,.
\end{align}
Applying the asymptotic expansion for Hankel function [Eq.~\eqref{eq:D_expansion}] at large argument, the second term of Eq.~\eqref{outside} 
yields the scattering amplitude $f(\phi)$ [Eq.~\eqref{outside_f}].

\section{Debye and ray series}\label{subsec:debye}

To derive the ray series, 
we first decompose $e^{2 i \delta_{j}}$ in Eq.~\eqref{eq:S_s} into the Debye series\cite{grandy2000}
\begin{align}
e^{2i\delta_{j}} &= R_{22}+\sum_{p=1}^{\infty} T_{21}T_{12}(R_{11})^{p - 1}\,, \label{eq:Debye_series}
\end{align}
where
\begin{align}
R_{22}&=\dfrac{ H_{j+1/2}^{(2)}(X)H_{j-1/2}^{(2)}(X')- H_{j-1/2}^{(2)}(X)H_{j+1/2}^{(2)}(X')}{d_{j}}\,,\\
T_{21}&=\dfrac{ H_{j-1/2}^{(2)}(X)H_{j+1/2}^{(1)}(X)- H_{j+1/2}^{(2)}(X)H_{j-1/2}^{(1)}(X)}{-d_{j}}\,,\\
d_{j}&\equiv H_{j-1/2}^{(1)}(X)H_{j+1/2}^{(2)}(X')- H_{j+1/2}^{(1)}(X)H_{j-1/2}^{(2)}(X')\,.
\end{align}
Coefficient $R_{11}$ ($T_{12}$) is obtained from $R_{22}$ ($T_{12}$) by interchanging $1$ with $2$ and $X'$ with $X$.
These $R$'s and $T$'s should not be confused with the
plane-wave reflection and transmission coefficients such as $R_{\textrm{in}}$ and $T_{\textrm{in}}$ in Sec.~\ref{sec:ray_pic}.

The scattering amplitude Eq.~\eqref{outside_f} can be now written as
\begin{align}
f(\phi)& = -\dfrac{i}{\sqrt{2\pi k}}\sum_{p=0}^{\infty} \sum_{j} D_{p}e^{i(j-1/2)\phi}\,,
\end{align} 
with $D_{0}= R_{22}$, $D_1=T_{12}T_{21}-1$
and $D_{p}=T_{12}T_{21}R_{11}^{p-1}$ for $p>1$.
Using the Poisson summation formula,
we obtain
\begin{align}
\sum_{j} D_{p} e^{i(j-1/2)\phi} = \sum_{m=-\infty}^{\infty} \int_{-\infty}^{\infty} d\lambda D_{p}e^{i\lambda(\phi + 2m\pi)}\,,
\label{eq:d1}
\end{align}
with $\lambda=j-1/2$.
For  $|\lambda - X| > X^{1/3}$, we can use
the Debye expansion of the Hankel function,     
\begin{align}
&H_{\lambda}^{(1,2)}(X)\simeq\left(\dfrac{2}{\pi}\right)^{1/2}(X^{2}-\lambda^{2})^{-1/4}\notag\\
&\times\exp \left\{\pm i\left[\sqrt{X^{2}-\lambda^{2}}-\lambda\cos^{-1}(\lambda/X)-\pi/4\right]\right\}\,,
\label{eq:D_expansion}
\end{align}
valid for $|\lambda| < x$, to approximate $D_p$.
After some tedious algebra, we find
\begin{align}
D_{p} \simeq
\begin{cases}
C_1(\alpha) e^{2i\delta_{1}} - 1, & p = 1\,,\\
C_p(\alpha) e^{2i\delta_{p}},     & p \neq 1\,,
\end{cases}
\label{eq:D_c_app}
\end{align}
where $C_p(\alpha)$ is defined in Eq.~\eqref{eq:C} and $\delta_{p}$ is
given by
\begin{align}
\delta_{p} = -&\left[
X\cos\alpha - \lambda\left(\dfrac{\pi}{2} + \alpha\right)
-\dfrac{\pi}{4}
\right]\notag\\
\mbox{} + p &\left[
X'\cos\beta - \lambda\left(\dfrac{\pi}{2} + \beta\right)
-\dfrac{\pi}{4}
\right]\,,
\end{align}
with 
\begin{align}
\alpha = -\sin^{-1}\left(\dfrac{\lambda}{X}\right),
\quad
\beta = -\sin^{-1}\left(\dfrac{\lambda}{X'}\right)\,.
\label{eq:lambda_alpha}
\end{align}
It can be shown that $D_p$ becomes very small at $\lambda > X$ for all $p$.
(For $p = 1$, having $-1$ term in $D_1$ is essential for this property.)
Thus, the infinite integration limits in Eq.~\eqref{eq:d1} can be replaced
by $\pm X$.
The integrand has a saddle-point determined by the condition
\begin{align}
2\dfrac{d\delta_{p}}{d\lambda} + \phi + 2m\pi = 0\,.
\label{eq:sp_phi}
\end{align}
Applying the saddle-point approximation, we obtain
Eq.~\eqref{circle_ray_p}.


\end{document}